\documentclass[
superscriptaddress,
final,
reprint,
showkeys,
prb,
aps,
floatfix
]{revtex4-1}
\usepackage{nameref}
\usepackage[dvips]{graphicx}
\usepackage[utf8]{inputenc}
\usepackage{multirow}
\usepackage{amsmath,amssymb,latexsym,MnSymbol,bm}

\usepackage[table]{xcolor}
\usepackage{xcolor}
\usepackage[%
  colorlinks=true,
  urlcolor=blue,
  linkcolor=blue,
  citecolor=blue
]{hyperref}
\setcitestyle{super}

\newcommand{\icol}[1]{
  \left[\begin{smallmatrix}#1\end{smallmatrix}\right]
}

\usepackage{scalerel,stackengine}
\stackMath
\newcommand\reallywidehat[1]{%
\savestack{\tmpbox}{\stretchto{%
  \scaleto{%
    \scalerel*[\widthof{\ensuremath{#1}}]{\kern-.6pt\bigwedge\kern-.6pt}%
    {\rule[-\textheight/2]{1ex}{\textheight}}
  }{\textheight}%
}{0.5ex}}%
\stackon[1pt]{#1}{\tmpbox}%
}

\begin{document}

\title{Effects of changing population or density on urban carbon dioxide emissions}

\author{Haroldo V.\ Ribeiro}
\email{hvr@dfi.uem.br}
\affiliation{Departamento de F\'isica, Universidade Estadual de Maring\'a -- Maring\'a, PR 87020-900, Brazil}

\author{Diego Rybski}
\email{ca-dr@rybski.de}
\affiliation{Potsdam Institute for Climate Impact Research -- PIK, Member of Leibniz Association, P.O. Box 601203, 14412 Potsdam, Germany}
\affiliation{Feodor Lynen Research Fellow of the Alexander von Humboldt Foundation}

\author{J\"urgen P.\ Kropp}
\affiliation{Potsdam Institute for Climate Impact Research -- PIK, Member of Leibniz Association, P.O. Box 601203, 14412 Potsdam, Germany}
\affiliation{Institute for Environmental Science and Geography, University of Potsdam, 14476 Potsdam, Germany}


\begin{abstract}
The question of whether urbanization contributes to increasing carbon dioxide emissions has been mainly investigated via scaling relationships with population or population density. However, these approaches overlook the correlations between population and area, and ignore possible interactions between these quantities. Here, we propose a generalized framework that simultaneously considers the effects of population and area along with possible interactions between these urban metrics. Our results significantly improve the description of emissions and reveal the coupled role between population and density on emissions. These models show that variations in emissions associated with proportionate changes in population or density may not only depend on the magnitude of these changes but also on the initial values of these quantities. For US areas, the larger the city, the higher is the impact of changing its population or density on its emissions; but population changes always have a greater effect on emissions than population density.
\end{abstract}

\maketitle

\section*{Introduction}
Carbon dioxide (CO$_2$) emissions are considered one of the main causes of Earth's climate change~\cite{Schellnhuber2006}. 
Despite covering only 0.4\,\%-0.9\,\% of global land surfaces \cite{EschHHKRZDS2017}, urban areas are responsible for more than 70\% of such emissions~\cite{johansson2012global,seto2014human}. This fact assigns cities a central role in pursuing solutions and mitigation strategies for the global climate change problem. Because of that, researchers from several disciplines have investigated the effects of urbanization on CO$_2$ emissions~\cite{NewmanK1989,lariviere1999modelling,brown2009geography,FragkiasLSS2013,OliveiraAM2014,tamayao2014us,LoufB2014SciRep,jones2014spatial,mohajeri2015co,ye2015sustainable,Gudipudi2016,chang2017there,rybski2017cities,chen2017coupling,muller2017does,gudipudi2019efficient}. The question of whether urbanization promotes or mitigates climate change is ubiquitous among these works, and the approaches to probe such issues differ, but can be roughly organized into two groups. 

Underlying the first approach, there is the so-called urban scaling hypothesis~\cite{bettencourt2007growth,bettencourt2010unified}, which states that city emissions ($C$) are described by a power-law function of population size ($P$), that is, $C \sim P^\beta$, where $\beta$ is the urban scaling exponent or the scale-invariant elasticity. For CO$_2$ emissions in US, researchers have reported a 1:1 relationship ($\beta \approx1$, constant returns to scale) with the population of metropolitan areas~\cite{FragkiasLSS2013}, while the same quantity was found to scale superlinearly ($\beta=1.46$, increasing returns to scale) when defining the US cities as connected urban spaces~\cite{OliveiraAM2014}. When considering local air pollution, an exponent $\beta\approx3/4$ (decreasing returns to scale) was observed for US metropolitan areas~\cite{muller2017does}. There is also evidence supporting the idea that the scaling between emissions and population depends on the degree of economic development of the urban systems, with increasing returns to scale ($\beta>1$) observed for cities of developing countries and economy of scale ($\beta<1$) for developed ones~\cite{rybski2017cities}. The second approach is focused on the understanding how population density affects CO$_2$ emissions per capita~\cite{NewmanK1989,lariviere1999modelling,baur2013urban,jones2014spatial,ye2015sustainable,Gudipudi2016}, that is, to investigate the relationship between $C/P$ and $P/A$, where $A$ stands for the urban unit area. A recent work has proposed that CO$_2$ emissions per capita (related to buildings and on-road sectors) and population density are related via the power-law $C/P \sim (P/A)^\alpha$, with an exponent $\alpha\approx-0.8$ for the US urban areas~\cite{Gudipudi2016}. 

Although these two bodies of the urban CO$_2$ literature are strongly linked by the purpose of understanding how urbanization affects climate change, they have operated widely independent from each other, and their approaches are perceived as different issues. Researchers using urban scaling are assuming population size as the most relevant urban feature for describing CO$_2$ emissions, while those working with the per capita density scaling consider population density as the most significant covariate. Both approaches, however, have produced controversial results regarding the influence of population or population density on urban emissions (see, for instance, Refs.~\cite{brown2009geography,jones2014spatial} and Supplementary Table~\ref{stab:1}). Large part of these discrepancies can be attributed to different methodologies for estimating CO$_2$ emissions and defining the boundaries of urban areas, but also because both approaches ignore that population and area are correlated~\cite{Stewart179Suggested,BattyF2011} and the influence of a possible interconnected role between these quantities on urban emissions. 

Inspired by the economic theory of production functions~\cite{heathfield1987introduction}, we propose here a new approach for investigating emissions in urban areas that simultaneously considers the effects of population and area along with possible interactions between these urban metrics. We show that our models recover the two conventional approaches when ignoring the effects of urban area (urban scaling) or when assuming that the emissions display constant returns to scale with population and area (per capita density scaling). When compared with the two conventional approaches, our models provide a significantly better description for the emissions in US urban areas. These results confirm the predictive power of the interactions between population and area, which in turn have intriguing consequences about the effect of these quantities on urban emissions. Our approach indicates that emissions may display decreasing or increasing returns to scale with population and area depending on whether the product $P\times A$ exceeds a particular threshold. We further find that the impact of a proportionate change in the population and density of a city on its emissions increase with its area but always have decreasing returns to scale; moreover, changes in population always have more impact on emissions than changes in density.

\section*{Results}
\subsection*{Urban scaling and per capita density scaling of emissions}
We start by revisiting how population scaling and per capita density scaling approaches have been applied for investigating CO$_2$ emissions in urban areas. To do so, we used the same dataset reported by Gudipudi et al.~\cite{Gudipudi2016} which comprises CO$_2$ emissions (sum of on-road and building emissions) in US urban areas in the year 2000. As described in \hyperref[sec:data]{Methods}, this dataset is constructed by combining gridded data from different sources, and by applying the city clustering algorithm~\cite{rozenfeld2008laws} for defining the urban units. There are a total of 3285 urban units and for each one we have population ($P$ in raw counts), area ($A$ in km$^2$), and CO$_2$ emissions ($C$ in tonnes of CO$_2$).

\begin{figure*}[ht]
\centering
\includegraphics[width=1\linewidth]{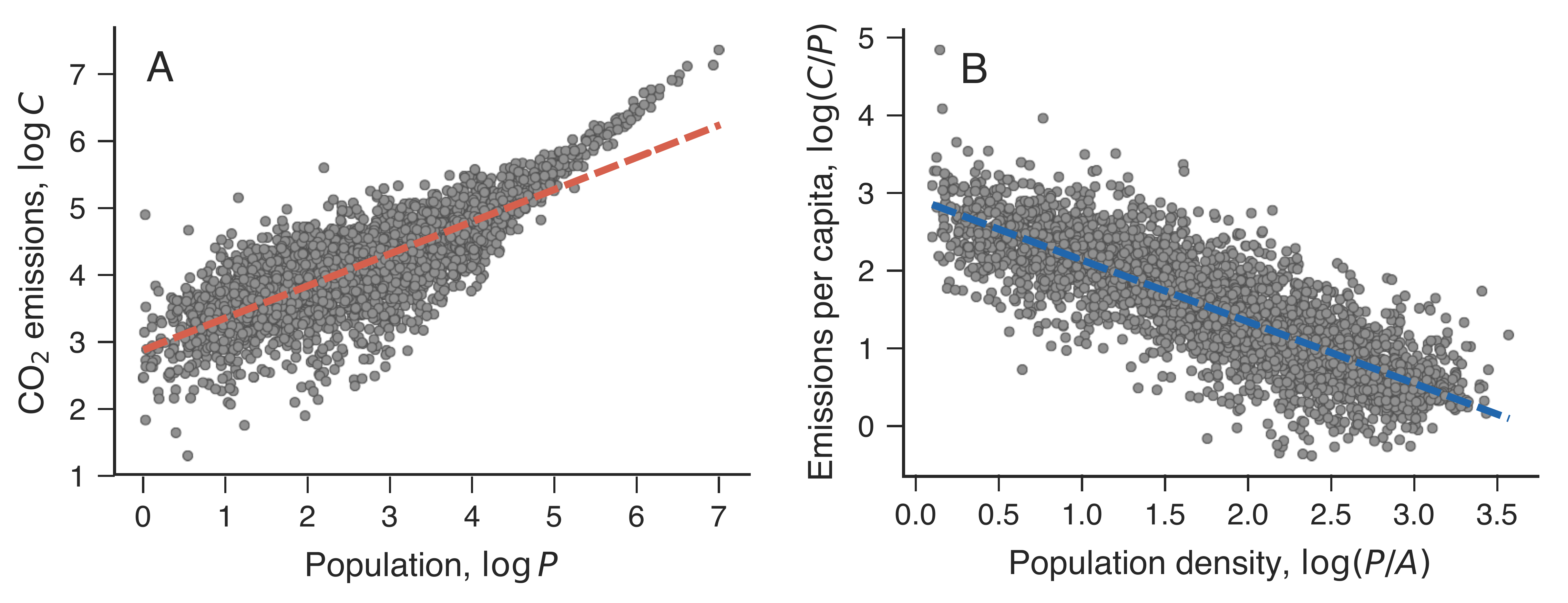}
\caption{{\bf Conventional approaches for investigating urban emissions.} (A) Urban scaling: the scaling relationship between CO$_2$ emissions ($C$) and population size ($P$). The dashed line represents a power-law fit [Eq.~(\ref{eq:urban_scaling})] with an exponent $\beta=0.48\pm0.01$. We observe that this model underestimates the emissions for large population sizes. (B) Per capita density scaling: the scaling law between CO$_2$ emissions per capita ($C/P$) and population density ($P/A$). The dashed line is a power-law fit [Eq.~(\ref{eq:density_scaling})] with an exponent $\alpha=-0.79\pm0.01$. In both plots, each dot is associated with a US urban unit obtained from the city clustering algorithm (\hyperref[sec:data]{Methods}) and all quantities are expressed in base-$10$ logarithmic scale. Emissions are measured in tonnes of CO$_2$, population in raw counts, and area in square kilometers.
}
\label{fig:1}
\end{figure*}

Having defined our variables, within the urban scaling framework, CO$_2$ emissions and population size are related via the power-law function
\begin{equation}\label{eq:urban_scaling}
C\sim P^{\beta}\,,
\end{equation}
where $\beta$ is the scaling exponent. To estimate the parameter $\beta$, we have applied the usual least-squares method to the relationship between $\log C$ and $\log P$. This approach leads to $\beta=0.48\pm0.01$ ($p$-value~$=0$, permutation test, Supplementary Figure~\ref{sfig:1}) and the relationship between both variables (on logarithm scale) is shown in Fig.~\ref{fig:1}A. At a cursory glance, this value of $\beta$ indicates a sublinear trend between emissions and population, so that a 1\% increase in the population level of a city associates with only 0.48\% increase in its CO$_2$ emissions. However, a closer inspection of Fig.~\ref{fig:1}A shows that Eq.~(\ref{eq:urban_scaling}) deviates systemically from the data and underestimates the emissions in large populated areas. In addition to that, the exponent $\beta$ is likely to be biased by the confounding effect of area because population and area of urban units are correlated to each other via a power-law relation~\cite{Stewart179Suggested,BattyF2011}.

On the other hand, within the per capita density scaling framework, the relationship between CO$_2$ emissions per capita and population density is described by the power-law function~\cite{Gudipudi2016}
\begin{equation}\label{eq:density_scaling}
C/P \sim (P/A)^{\alpha}\,,
\end{equation}
where $\alpha$ is another scaling exponent. Figure~\ref{fig:1}B illustrates this relationship on logarithmic scale ($\log C/P$ versus $\log P/A$) from which we have estimated $\alpha=-0.79\pm0.01$ ($p$-value~$=0$, permutation test, Supplementary Figure~\ref{sfig:1}) via ordinary-least-squares method. We observe that Eq.~(\ref{eq:density_scaling}) is slightly better than Eq.~(\ref{eq:urban_scaling}) for describing our data (Supplementary Figure~\ref{sfig:2}), and does not seem to exhibit any systematic deviation. Our estimate for $\alpha$ is in agreement with the results reported by Gudipudi et al.~\cite{Gudipudi2016}, and indicates that every $1\%$ increase in the population density of a city associates with a $0.79\%$ reduction in its CO$_2$ emissions per capita. However, and similarly to the urban scaling case, the exponent $\alpha$ is also likely to be biased by the confounding effect of population, since density and population values are also correlated~\cite{Stewart179Suggested,BattyF2011}.

In order to improve the description of urban emissions, we propose here an analogy with the economic theory of production functions. This theory has a central role in several branches of economics~\cite{heathfield1987introduction}, and in general terms, a production function establishes a mathematical relationship between inputs (capital, labor, land, etc.) and output of goods (iron, cars, wheat, etc.) in some production process. Using this mathematical description, economists may ask how much output can be produced with particular combinations of inputs, and what are the alternatives (in terms of inputs) for producing a particular good. These ideas transpose well into our context assuming CO$_2$ emissions as the output and population and area as the inputs in a production process mediated by cities. Similarly to a two-factor production model, we thus consider that $C = F(P,A)$, where $F(\dots)$ stands for the form of the production function. By putting this analogy forward, we establish a more general approach for modeling urban emissions that simultaneously accounts for the effects of population and area along with possible interactions between these urban metrics. As we shall verify, we can borrow not only functional forms from the theory of production functions but also key concepts that are very useful in the context of urban emissions (\hyperref[sec:analogy]{Methods}).

\subsection*{Cobb-Douglas model of urban carbon emissions}
We start this analogy with the Cobb-Douglas model~\cite{CobbD1928}, arguably the most widely known and used production function~\cite{heathfield1987introduction}. In our case, it takes the form
\begin{equation}\label{eq:cobb_douglas_scaling}
C \sim P^{\beta_P}\,A^{\beta_A}~~\text{or}~~\log C \sim \beta_P \log P + \beta_A \log A\,,
\end{equation}
where $\beta_P$ and $\beta_A$ are two independent exponents. We immediately notice that this model recovers the urban scaling [Eq.~(\ref{eq:urban_scaling})] if $\beta_P=\beta$ and $\beta_A=0$ (that is, when ignoring the effect of area) and the per capita density scaling [Eq.~(\ref{eq:density_scaling})] if $\beta_P=\alpha+1$ and $\beta_A=-\alpha$. We further remark that the Cobb-Douglas model can be obtained from Eqs.~(\ref{eq:urban_scaling}) and (\ref{eq:density_scaling}) if we consider the empirical relation between population and area~\cite{Stewart179Suggested,BattyF2011} (\hyperref[sec:connecting_cobb]{Methods}). Similarly to the models of Eqs.~(\ref{eq:urban_scaling}) and (\ref{eq:density_scaling}), the Cobb-Douglas function exhibits a scale-invariant elasticity $\varepsilon=\beta_P+\beta_A$ (\hyperref[sec:analogy]{Methods}), meaning that a proportionate increase in emissions associated with a proportionate increase in population and area is independent of $P$ and $A$. Thus, when $\beta_P+\beta_A<1$ there are decreasing returns to scale (doubling $P$ and $A$ implies less than doubling $C$), whereas if $\beta_P+\beta_A>1$ there are increasing returns to scale (doubling $P$ and $A$ implies more than doubling $C$), and only for $\beta_P+\beta_A=1$ this model presents constant returns to scale (doubling $P$ and $A$ implies exactly doubling $C$). Thus, we notice that Eq.~(\ref{eq:density_scaling}) is a particular case of the Cobb-Douglas model with constant returns to scale. On the other hand, without any constraint for the exponents $\beta_P$ and $\beta_A$, the Cobb-Douglas model represents a genuine generalization that cannot be related to Eqs.~(\ref{eq:urban_scaling}) and (\ref{eq:density_scaling}). In addition to that, we can interpret Eq.~(\ref{eq:cobb_douglas_scaling}) as the result of accounting for the confounding effect of area $A$ within the urban scaling framework [Eq.~(\ref{eq:urban_scaling})] via a multiple linear regression (in log-transformed variables).

\begin{figure*}[ht]
\centering
\includegraphics[width=\linewidth]{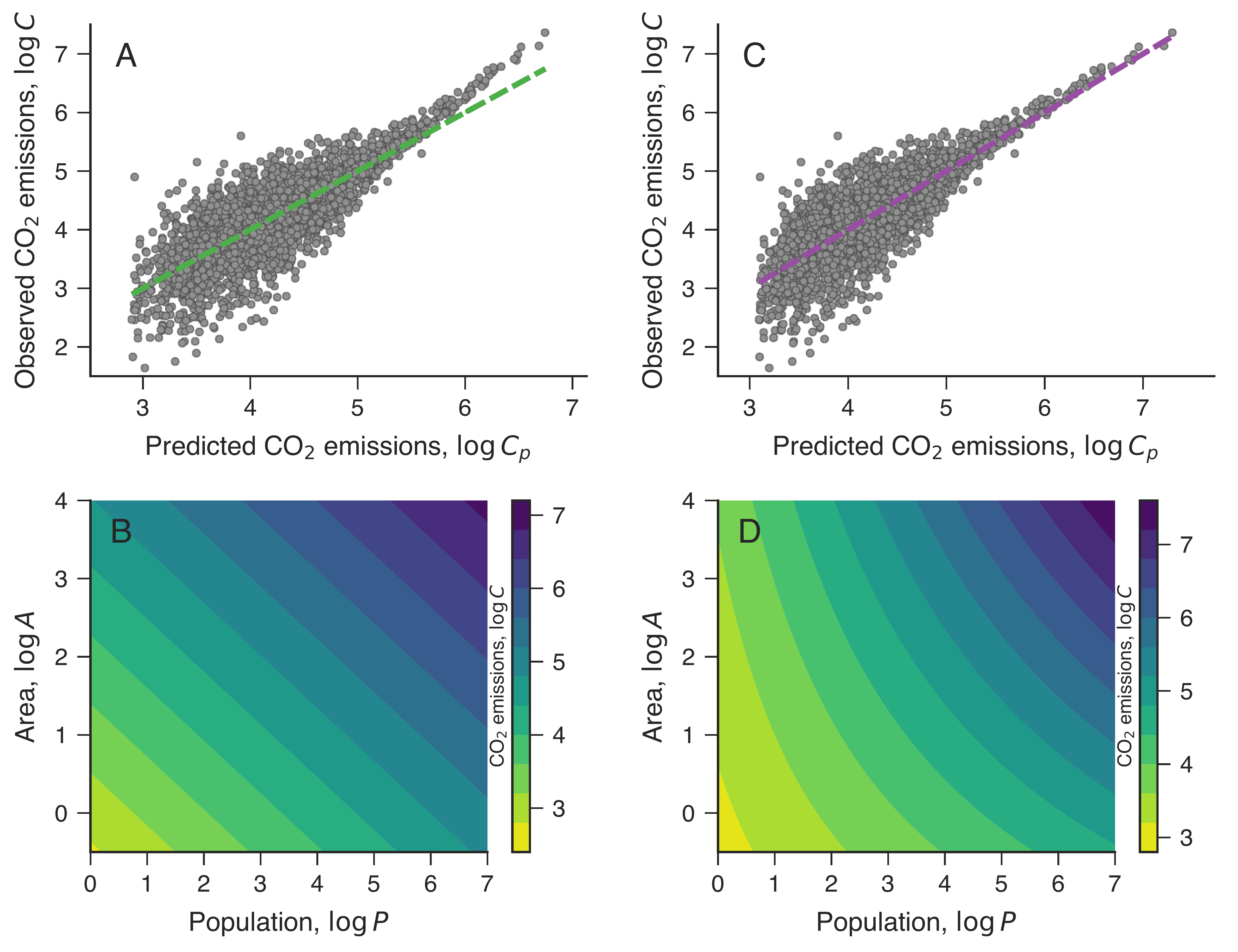}
\caption{{\bf The interplay between population and area on CO$_2$ emissions.} (A) Scatter plot of the observed values of CO$_2$ emissions ($C$) and those predicted ($C_P$) by the Cobb-Douglas model [Eq.~(\ref{eq:cobb_douglas_scaling}) with $\beta_P=0.31\pm0.01$ and $\beta_A=0.45\pm0.03$]. This model is a significantly better fit when compared with the urban scaling and the per capita density scaling models (Supplementary Figure~\ref{sfig:2}). (B) A contour plot of Eq.~(\ref{eq:cobb_douglas_scaling}) as a function of $P$ and $A$ on logarithmic scale. The straight isolines/isoquants show how population and area must change in order to keep the emissions unchanged. (C) Scatter plot between the observed and predicted CO$_2$ emissions obtained from the translog model [Eq.~(\ref{eq:translog_scaling}) with $\beta_P=0.28\pm0.02$, $\beta_A=0.14\pm0.05$, and $\beta_C=0.07\pm0.01$]. This model further refines the goodness of the predictions (Supplementary Figure~\ref{sfig:2}), particularly reducing the bias in urban areas with high emissions. (D) A contour plot of Eq.~(\ref{eq:translog_scaling}) as a function of $P$ and $A$. We note that the isolines/isoquants of this model are not straight lines as those from the Cobb-Douglas model. We have employed base-$10$ logarithmic quantities in all panels.
}
\label{fig:2}
\end{figure*}

Although the model of Eq.~(\ref{eq:cobb_douglas_scaling}) may represent a better description for CO$_2$ emissions, it introduces some drawbacks related to the use of ordinary-least-squares for finding the best fitting parameters $\beta_P$ and $\beta_A$. This happens because population and area are correlated to each other, a problem known as multicollinearity and that can lead to unstable estimates for the model parameters. As detailed in \hyperref[sec:fitting]{Methods}, we have applied the regularization approach of the ridge regression~\cite{hoerl1970ridge,hastie_elements_2016} in order to account for this problem. To state briefly, the ridge regression adds a penalty/regularization term proportional to the square of the magnitude of coefficients upon the residual sum of squares, which in turn stabilizes the regression coefficients and accounts for the multicollinearity. This approach yields ${\beta}_P=0.31\pm0.01$ and ${\beta}_A=0.45\pm0.03$ ($p$-values~$=0$, permutation test, Supplementary Figure~\ref{sfig:3}), and Fig.~\ref{fig:2}A shows the relationship between the actual values of the CO$_2$ emissions and those predicted by Eq.~(\ref{eq:cobb_douglas_scaling}). We have verified that the Cobb-Douglas model provides a significantly better fit to our data (Supplementary Figure~\ref{sfig:2}) when compared with the urban scaling [Eq.~(\ref{eq:urban_scaling})] and the per capita density scaling approaches [Eq.~(\ref{eq:density_scaling})]. Moreover, the fact that ${\beta}_P$ is much smaller than $\beta$ reinforces the idea that the urban scaling approach is indeed affected by the confounding effect of the area.

Because $\beta_P + \beta_A<1$, our results indicate that CO$_2$ emissions display diminishing returns as population and area are incrementally increased by the same factor (that is, keeping density constant). In particular, our estimates indicate that every $1$\% increase in both the population and area of a city associates with a $0.76$\% increase in its emissions. The interconnected role of population and area on CO$_2$ emissions is better visualized in Fig.~\ref{fig:2}B, where we depict a contour plot of Eq.~(\ref{eq:cobb_douglas_scaling}) on logarithmic scale. In this representation, the isoquants (or isolines) are described by straight lines [$\log A\sim -(\beta_P/\beta_A) \log P$] and show how population and area must change to keep emissions constant. These isoquants also indicate that if the population of a city increases, its population density must also increase to keep emissions unaltered. This behavior is better understood by rewriting Eq.~(\ref{eq:cobb_douglas_scaling}) as $C \sim P^{\beta_P+\beta_A}\,(P/A)^{-\beta_A}$ and noticing that $\beta_P>0$ and $\beta_A>0$ (Supplementary Figure~\ref{sfig:4} shows the contour plot in terms of population density). From this form of Eq.~(\ref{eq:cobb_douglas_scaling}), we also conclude that a proportionate change in population has more impact on CO$_2$ emissions than a proportionate change in density since $|\beta_P+\beta_A|>|\beta_A|$. For a particular urban unit in our data, this means that if its population decreases by 1\% while its density remains constant (that is, a 1.01\% increase in its area), the model predicts a 0.76\% reduction in its CO$_2$ emissions; whereas a 1\% raise in its density while population is unaltered (that is, a 1\% reduction in its area), implies only 0.45\% decrease in its CO$_2$ emissions. 

Another interesting aspect of an isoquant is its slope, a quantity known as the technical rate of substitution in economics~\cite{heathfield1987introduction} (\hyperref[sec:analogy]{Methods}). In our context, this slope measures how much the population of a city should change in response to alterations in its area in order to keep the same level of emissions. For Eq.~(\ref{eq:cobb_douglas_scaling}), the slopes of the isoquants are $\frac{dA}{dP} = -\frac{\beta_P}{\beta_A(P/A)}$, and thus, they are completely determined by the city density (assuming that $\beta_P$ and $\beta_A$ are known). If we consider a logarithmic scale, these slopes are equal to $\frac{d\log A}{d \log P} = -\beta_P/\beta_A$ regardless of the values of $P$ and $A$ (as we see in Fig.~\ref{fig:2}B). In economics, these isoquants are also analyzed in terms of the so-called elasticity of substitution~\cite{heathfield1987introduction} (\hyperref[sec:analogy]{Methods}), a dimensionless measure that (mapped to our case) quantifies the efficiency at which population and area substitute each other, and that somehow reflects the shape of the isoquants~\cite{heathfield1987introduction}. Usually, more L-shaped isoquants are associated with low elasticity of substitution (that is, there is no room for replacing $A$ by $P$ while keeping emissions constant), whereas more linear/smooth isoquants tend to have high elasticity of substitution (it is easy to replace $A$ by $P$ while keeping emissions constant). The Cobb-Douglas model has unitary elasticity of substitution, regardless of the values of $P$, $A$, and $C$, and also the exponents $\beta_P$ and $\beta_A$~\cite{heathfield1987introduction}. Thus, although the Cobb-Douglas model provides a better fit to our data [compared with Eqs. (\ref{eq:urban_scaling}) and (\ref{eq:density_scaling})], it also makes a series of assumptions that do not have any compelling reasons to hold true in urban systems (as it also happens in economics~\cite{heathfield1987introduction}). Moreover, Fig.~\ref{fig:2}A shows that Eq.~(\ref{eq:cobb_douglas_scaling}) has a bias for large values of $C$, which indicates that relaxing some underlying assumptions of the Cobb-Douglas model may lead to a better description of the emissions.

We first relax the condition of unitary elasticity of substitution by considering a model based on the constant elasticity of substitution (CES) production function~\cite{arrow1961capital,heathfield1987introduction}
\begin{equation}\label{eq:CES_scaling}
C \sim (\beta_P P^{-\gamma} + \beta_A A^{-\gamma})^{-1/\gamma}\,,
\end{equation}
where $\gamma$ is a parameter and $\beta_P+\beta_A=1$. The CES model emerged as a generalization of the Cobb-Douglas function (which is recovered when $\gamma\to0$) exactly because economists have considered the assumption of unitary elasticity of substitution as unduly restrictive~\cite{arrow1961capital}. As the name suggests, the CES model has a constant elasticity of substitution equal to $1/(\gamma+1)$, and thus, by varying $\gamma$ we have a wide range of possible elasticities. We have adjusted Eq.~(\ref{eq:CES_scaling}) to our data using the Levenberg-Marquardt algorithm. However, these fits are very problematic because of the non-linear nature of the model. Depending on the initial guess used for the parameters, we find very unstable and meaningless results. These fits also lead to very large variations in the parameters when applying a resampling strategy to our data. Furthermore, we have observed that even when the fits converge, the CES model does not represent the best description to our data. In addition to the non-linearity, this happens because the CES model also carries crucial assumptions that appear not to hold in our case. This model has unitary elasticity of scale ($\varepsilon=1$, that is, it assumes that by doubling $P$ and $A$ implies doubling $C$), and similarly to Cobb-Douglas, the slopes of the isoquants $\left(\frac{dA}{dP} = -\frac{\beta_P}{\beta_A(P/A)^{\gamma+1}}\right)$ are determined solely by the city density.

\subsection*{Transcendental logarithm model of emissions}
To overcome these limitations and constraints, we considered a more general production function known as the transcendental logarithm (translog) model~\cite{christensen1973transcendental,heathfield1987introduction}. Just like the CES is a natural extension of the Cobb-Douglas function, the translog model represents the next logical step towards a more flexible function for modeling the CO$_2$ emissions in terms of $P$ and $A$. This model is written as
\begin{equation}\label{eq:translog_scaling}
\log{C} \sim \beta_P \log P + \beta_A \log A + \beta_C \log (P) \log (A)\,,
\end{equation}
where $\beta_C$ is an additional parameter. Differently from Eqs.~(\ref{eq:cobb_douglas_scaling}) and (\ref{eq:CES_scaling}), the translog model has a non-constant elasticity of scale [$\varepsilon = \beta_P + \beta_A + \beta_C \log(P A)$], meaning that proportionate changes in the emissions associated with proportionate changes in population and area depend on the initial values of $P$ and $A$. Furthermore, the elasticity of substitution of Eq.~(\ref{eq:translog_scaling}) varies with $P$, $A$ and $C$, and the isoquant slope depends on $P$ and $A$ (\hyperref[sec:analogy]{Methods}). By comparing the translog model with the Cobb-Douglas, we notice that all this additional flexibility is solely related to the inclusion of the interaction term [$\beta_C \log (P) \log (A)$] between population and area, where $\beta_C$ quantifies the intensity of this interaction. It is this interaction term that allows the effect of population and area on emissions to vary with $P$ and $A$. This effect is better understood in terms of the marginal products~\cite{heathfield1987introduction} (\hyperref[sec:analogy]{Methods}), an economic quantity that mapped to our context represents the response in CO$_2$ emissions caused by changes in population or area. The marginal product of population (in logarithmic scale) is defined as $\frac{d\log C}{d\log P}=\beta_P +\beta_C \log A$, while the marginal product of area is $\frac{d\log C}{d\log A}=\beta_A +\beta_C \log P$. Thus, for instance, we observe that the marginal product of population depends on the area of the urban unit. This behavior contrasts with the Cobb-Douglas predictions (corresponding to $\beta_C=0$), in which the marginal products are independent of $P$ and $A$.

As in the Cobb-Douglas case, we have considered the ridge regression approach in order to account for the multicollinearity and adjust Eq.~(\ref{eq:translog_scaling}) to our data (\hyperref[sec:fitting]{Methods}). This approach yields $\beta_P=0.28\pm 0.02$, $\beta_A = 0.14\pm0.05$, and $\beta_C=0.07\pm0.01$ ($p$-values~$=0$, permutation test, Supplementary Figure~\ref{sfig:3}). We further note that the translog model is a significantly better fit to our data when compared with the previously-discussed models (Supplementary Figure~\ref{sfig:2}). This fact is highlighted in Fig.~\ref{fig:2}C, where we observe that the model of Eq.~(\ref{eq:translog_scaling}) refines the quality of the predictions and reduces the bias observed for Eq.~(\ref{eq:cobb_douglas_scaling}) in urban areas with high emissions. We further remark that all fitted parameters of Eq.~(\ref{eq:translog_scaling}) are significantly different from zero, confirming the predictive power of the interaction term between population and area. Therefore, in addition to improving the description of the CO$_2$ emissions, Eq.~(\ref{eq:translog_scaling}) reveals that the effect of population and area on the emissions intensifies with the increase of urban population and area. This becomes clear by noticing that the elasticity of scale increases with $P$ and $A$ [$\varepsilon = 0.42 + 0.07 \log (P A)$], that is, the more populous and the more widespread a city is, the larger is the impact of a proportionate change in its population and area (a growth with constant density) on its emissions. It is further intriguing to notice that, because the elasticity of scale varies with $P$ and $A$, the translog model displays decreasing, increasing, or constant returns to scale depending on whether the product $\Omega = P A$ is respectively smaller, larger or equal to the critical value  
\begin{equation}
\Omega^* = 10^{\frac{1-\beta_P-\beta_A}{\beta_C}}\,.
\end{equation}
For the US data $\Omega^*\approx 1.93\times10^8$, and thus, cities having $\Omega<\Omega^*$ display decreasing returns to scale, while those having $\Omega>\Omega^*$ feature increasing returns to scale. For instance, the translog model predicts that a 1\% increase in population and area of a large city with $P=8\times10^6$ and $A=6,000$~km$^2$ (roughly the size of Chicago, $\Omega =4.8\times10^{10}$) associates with 1.17\% raise in its emissions, whereas the same change in a relatively small city with $P=90,000$ and $A=140$~km$^2$ (roughly the size of Santa Fe, $\Omega =12.6\times10^{6}$) associates with only 0.92\% raise in its emissions. 

The interconnected role of population and area on CO$_2$ emissions is better visualized in the contour plot of Eq.~(\ref{eq:translog_scaling}) shown in Fig.~\ref{fig:2}D. In comparison with Fig.~\ref{fig:2}B, we note that the interaction term bends the isoquants upward and make their slopes $\left[\frac{dA}{dP} = \frac{-1}{(P/A)}\left(\frac{\beta_P +\beta_C \log A}{\beta_A +\beta_C \log P}\right)\right]$ a function of $P$ and $A$, and not only of the city density as in the Cobb-Douglas model. We further observe that the spacing between the isoquants representing equally incremented values of $\log C$ changes with the values of $\log P$ and $\log A$. This behavior contrasts with the equally spaced isoquants produced by the model of Eq.~(\ref{eq:cobb_douglas_scaling}), and emphasizes that proportionate changes in emissions caused by changes in both population and area depend not only on the intensity of the changes but also on the initial values of population and area. 

In terms of density, we can rewrite Eq.~(\ref{eq:translog_scaling}) as 
\begin{equation}\label{eq:translog_scaling_density}
\begin{split}
C &\sim P^{\beta_P + \beta_A + \beta_C\log P}(P/A)^{-\beta_A-\beta_C\log P}\,,
\end{split}
\end{equation}
and from this expression, we find that the elasticity of scale in terms of population and density is $\varepsilon=\beta_P + \beta_C \log A$. Thus, the more widespread a city is, the larger is the impact of a proportionate change of population and density (that is, area remains constant) on its emissions. In a concrete example, our estimates indicate that a $1\%$ raise in population and density of a large city with $6,000$~km$^2$ associates with a $0.54\%$ increase in its emissions, while the same change in a city with $140$~km$^2$ correlates with a $0.42\%$ raise in its emissions. Furthermore, in terms of population and density, increasing returns to scale ($\varepsilon>1$) is only possible for urban areas exceeding the critical value $A^{*}=10^{(1-\beta_P)/\beta_C}$. For the US data, $A^{*}\approx1.93\times10^{10}$~km$ ^2$, an area roughly corresponding to 38 times the area of Earth. Therefore, our estimates show that only decreasing returns to scale are possible when the CO$_2$ emissions are described 
in terms of population and density. 

The translog model of Eq.~(\ref{eq:translog_scaling_density}) also allows us to verify whether changes in population have more impact on the emissions than changes in density. To do so, we have compared the absolute values of the marginal products of population $\left[\frac{d\log C}{d\log P} = \beta_P +\beta_A +\beta_C \log (P A)\right]$ and density $\left[\frac{d\log C}{d\log (P/A)}=-\beta_A -\beta_C \log P\right]$. The marginal product of population represents the response in emissions associated with changes in population when density remains constant, whereas the marginal product of density expresses the response in emissions caused by a change in density when population remains constant. The inequality $\left|\frac{d\log C}{d\log P}\right|>\left|\frac{d\log C}{d\log (P/A)}\right|$ simplifies to $A>A^{*}=10^{-\beta_P/\beta_C}$ when $\beta_P,\beta_A,\beta_C>0$ and $A,P>1$ (assumptions that agree with our estimates). By plugging the estimated values of $\beta_P$ and $\beta_C$, this condition becomes $A>10^{-4}$~km$^2$, and hence we conclude that population size always have more impact on emissions than changes in population density. We have reached the same conclusion with the simpler model of Eq.~(\ref{eq:cobb_douglas_scaling}). However, and as we have verified, the translog approach further refines the description of CO$_2$ emissions and indicates that the impact of population and density on the emissions changes with the population and density of the cities. For instance, according to our estimates for the US data, a 1\% increase in the density of a town with 10,000 inhabitants associates with a 0.42\% reduction in its emissions, while the same change in a larger city with 1 million inhabitants associates with a 0.56\% reduction in its emissions. This behavior contrasts with results of Eq.~(\ref{eq:cobb_douglas_scaling}), whose predictions related to changes in the density of a city are independent of its population. By carrying this predictions forward and in line with other studies~\cite{dodman2011forces}, our results suggest that the densification of large populated urban areas is likely to have important contributions to the reduction of urban CO$_2$ emissions.

\section*{Discussion}
We have shown that two conventional approaches used to study the effect of urbanization on urban CO$_2$ emissions suffer from confounding effects, and are unable to describe the interconnected role of population and area on urban emissions. Inspired by the economic theory of production functions, we have proposed new models for describing urban emissions simultaneously in terms of the population and area (or population density) of urban units. These models not only account for such confounding factors but significantly refine the description of the emissions in terms of urban quantities. In addition to being better fits to data, our models reveal intriguing aspects about the interplay between population and area (or density) on urban emissions that would be entirely neglected under the urban scaling or the per capita density scaling frameworks. Among these findings, our results indicate that variations in emissions associated with proportionate changes in population and area do not only depend on the magnitude of these changes but also on the product $\Omega=PA$. In particular, depending on whether $\Omega$ exceeds or not the critical value $\Omega^*$, urban emissions can display increasing ($\Omega>\Omega^*$) or decreasing returns to scale ($\Omega<\Omega^*$) with population and area. When described in terms of population and density, we have found that urban emissions display decreasing returns to scale, meaning that doubling population and density of a city always associates with less than doubling its emissions. We have further verified that changing the population of a city has more impact on its emissions than changing its density. In spite of that and in general terms, our models define conditions under which changes in population have more impact than changes in density (or vice versa) on emissions and further predict a transition-like behavior where the dominant role changes between these quantities if the urban area exceeds a threshold value ($A^*$).  

Our work has however its limitations in a sense that ideally the comparison between the effects of population and area (or density) on the emissions should be made after accounting for every other factor (such as economic activity, technology, and even individual attitudes) that possibly affects urban emissions. Thus, while our models account for the confounding effects of area (and density), the emissions may also be affected by other confounding factors not available in our dataset. One possibility for addressing this problem would be to include further control variables in our models, an approach that somehow resembles the IPAT equations~\cite{dietz1997effects,Chertow_IPAT_2000,Waggoner2002framework}, a framework proposed to model environmental impact (I) as the product of population (P), affluence (A), and technology (T), but with the advantage of considering population density (or area) as a predictor and allowing the interactions among such factors. Another possibility for overcoming these possible confounding effects is to combine our approach with the recently proposed urban Kaya scaling~\cite{gudipudi2019efficient} that relates CO$_2$ emissions, population, gross domestic product, and energy consumption. Combining these different approaches into a single and coherent framework could represent an exciting perspective for solving the economics of urban CO$_2$ emissions and defining its most important covariates. However, such endeavors require homogeneous and consistent data, which are still scarce on large spatial scales. While moving from urban units defined in terms of connected urban spaces to some political or administrative divisions would be a possibility, this approach is likely to introduce serious bias to the empirical estimates~\cite{ArcauteHFYJB2014} in addition to overestimating urban areas~\cite{cai2014urban} (\hyperref[sec:data]{Methods}). Another important limitation of our study is related to the intra-city processes and urban characteristics that cannot be accounted for only by population and area (or density). Case studies on this subject have shown that the urban form and intra-city population distribution have a substantial impact on urban emissions, particularly on transportation emissions. Cities from rapid developing countries such as China and India have undergone through a remarkable decentralization and suburban growth processes~\cite{wang2017urban}. These more dispersed urban forms and the consequent increase of the population living in urban frontier areas have direct implications for commuting and contribute to increasing CO$_2$ emissions~\cite{wang2014changing,wang2017urban,yang2019urban}. Regarding this aspect, it would be very interesting for future works to include possible covariates able to account for population imbalance and urban form in our models and thus quantify their impact in a large scale study. 
\vspace{-0.13cm}

Despite of these limitations, our work adds to the current understanding about the role of urbanization on CO$_2$ emissions, 
shedding light particularly upon the role of population and urban area (including their interactions) on urban emissions. Such interactions are completely overlooked within the urban scaling and per capita density scaling approaches and our work demonstrates that they play an important role on the description of urban emissions. Finally, our framework can be directly applied to other urban metrics in the place of emissions, opening thus a considerable range of possibilities for investigating the interplay between population and area (or population and density) over other important urban metrics.

\section*{Methods}

\subsection*{Dataset}\label{sec:data}
Our dataset is the same as analyzed by Gudipudi et al.~\cite{Gudipudi2016} and comprises the CO$_2$ emissions from urban areas of the United States (US) in the year 2000. As described by Gudipudi et al., this dataset is compiled from different sources through the following steps. Firstly, gridded population data are obtained from the Global Rural-Urban Mapping Project (GRUMP)~\cite{GRUMP} and the Global Land Cover Dataset (GLC)~\cite{GLC}. Both datasets are from the year 2000 and are available at a spatial grid resolution of 1\,km $\times$ 1\,km. The GRUMP data are spatially overlaid to the GLC in order to attribute population to the land use, which is classified as urban and non-urban. Next, sectoral emissions data (building and transportation) are obtained from the Vulcan Project~\cite{gurney2010vulcan}. This dataset is initially available at a resolution of 10\,km $\times$ 10\,km and has been down-scaled to 1\,km $\times$ 1\,km to be superimposed on the populated settlements. This process consists of equally splitting the emissions located in a cell of the Vulcan project among all overlapping population cells classified as urban. Finally, the city clustering algorithm (CCA)~\cite{rozenfeld2008laws} is used to systematically define the urban units, leading to the population size ($P$ in raw counts), area ($A$ in square kilometers), and CO$_2$ emissions ($C$ in tonnes of CO$_2$) for each urban unit. CCA is an iterative clustering algorithm that assigns any two cells to the same cluster if their distance is smaller or equal than a predefined threshold distance $l$. 

All results presented in our manuscript have been obtained using $l=5$\,km, a threshold distance that does not overestimate nor underestimate the urban extents~\cite{Gudipudi2016}. However, we have verified that our conclusions are very robust against variations in $l$ from $l=1$\,km to $l=10$\,km. In particular, the translog model [Eq.~(\ref{eq:translog_scaling})] always provides the best description for the US data (Supplementary Figure~\ref{sfig:5}). We have observed that the scaling exponents $\beta$ and $\alpha$ present slightly variations with threshold distance $l$ (Supplementary Figure~\ref{sfig:6}) that have no implications for the conclusions drawn from our findings. The parameters of the Cobb-Douglas ($\beta_P$ and $\beta_A$) and translog ($\beta_P$, $\beta_A$, and $\beta_C$) moldels present somewhat larger variations (Supplementary Figure~\ref{sfig:7}), but all remain statistically significant, particularly the parameter related to the interaction term in the translog model ($\beta_C$). These changes affect our point estimates for the Cobb-Douglas (Supplementary Figure~\ref{sfig:8}) and translog (Supplementary Figure~\ref{sfig:9}) models. In general, changes in emissions associated with a 1\% increase in $P$ and $A$ tend to increase with $l$. Similarly, the reduction in emissions associated with -1\% change in $P$ with fixed density or with a 1\% change in density with fixed $P$ also increase with $l$. This dependence on $l$ is smaller in the translog than in the Cobb-Douglas model. The critical product $\Omega^*$ displays a decreasing trend with $l$ (Supplementary Figure~\ref{sfig:10}), that in turn affects the point from which the effect of $P$ and $A$ on $C$ changes from decreasing to increasing returns to scale. The critical value $A^*$ is approximately independent of $l$ (Supplementary Figure~\ref{sfig:10}), indicating that changes in population have more impact than changes in density on emissions regardless the value of $l$. We have further verified that the contour plots of the translog function [Eq.~(\ref{eq:translog_scaling})] are very similar for different values of $l$ (Supplementary Figure~\ref{sfig:11}). 

These robustness tests are important because there has been a great debate about how to accurately define the correct boundaries of a city~\cite{ArcauteHFYJB2014}. In spite of that, there is still no consensus on this issue nor has a fail-safe procedure for defining the correct boundaries of a city been proposed yet. This issue also has great similarity with the more general concept of clustering and a quite similar issue arises when applying community detection algorithms in complex networks. All these topics have been exhaustively studied, but no silver bullet method exists. In the case of cities, additional complexity emerges because some urban indicators are more spatially constrained than others, and also because people commute to work and move from place to place in the long run.

Partly because of the seminal works by Bettencourt et al.~\cite{bettencourt2007growth}, the use of functional definitions for cities such as the Metropolitan Statistical Areas (MSAs) in US or larger urban units (LUZ) and metropolitan areas (MAs) in Europe have become very popular. These definitions are based on the idea of integrated socio-economic units and appear to be the gold standard for the urban scaling hypothesis as well as other purposes. In particular, MSAs are defined by a core county (or even more than one) having at least 50,000 people aggregated with adjacent counties that display a high degree of interaction (social and economic) with the central county (as measured by commuting flows). While this definition may work well for studying urban scaling, it is very problematic in our case. Since MSAs are made up of counties, they often include vast rural areas which in turn hugely overestimate the urban extent areas. In addition to that, MSAs can also fragment urban clusters into different pieces.

These problems are the main reason why we have chosen the CCA to define the urban units in our study. It is worth noticing that the CO$_2$ emissions we have analyzed are from building and transportation, and thus primarily associated with settlements where people reside and commute. Because of that, we argue that the approach of combining gridded data from the Vulcan project, GLC and GRUMP with CCA (proposed in Ref.~\cite{Gudipudi2016}) provides more precise emission estimates and associated spatial extents of urban clusters. Despite these problems and limitations, we have also applied our models to emissions data associated with MSAs. To do so, we have used the dataset provided by Fragkias et al.~\cite{FragkiasLSS2013} and considered the emissions from the year 2000 (the same used in our analysis). Indeed, these data allowed us to consider not only MSAs but also Micropolitan Areas ($\mu$SAs are defined as labor market areas with population between 10,000 and 50,000 people that are also made up of counties) and both together (Core Based Statistical Areas -- CBSAs). 

Supplementary Figure~\ref{sfig:12} shows the urban and per capita density scaling laws for MSAs. We notice that the quality of these relationships is not comparable with those reported in Fig.~\ref{fig:1}, and a similar situation happens for $\mu$SAs and CBSAs. Supplementary Figure~\ref{sfig:13}A compares the scaling exponents obtained from MSAs, $\mu$SAs, and CBSAs with those obtained via CCA with $l=5$~km. We observe that the values of $\beta$ estimated from these functional city definitions are much closer to one (in agreement with Fragkias et al.~\cite{FragkiasLSS2013}) than the values obtained with the CCA approach. As discussed in more detail by Bettencourt et al.~\cite{bettencourt2013hypothesis}, the disaggregation from the true urban unit can either introduce a positive or negative bias in the estimates of $\beta$. On the other hand, the aggregation of different true urban units tends to make $\beta$ closer to one. In the case of MSAs, it is likely that both disaggregation and aggregation effects play some role, but the fact that $\beta$ is smaller for $\mu$SAs than MSAs suggests that aggregation may have greater influence. Due to the poor quality of the per capita density scaling, it is hard to directly compare the values of $\alpha$ obtained from MSAs, $\mu$SAs and CBSAs with those estimated with the CCA; however, Supplementary Figure~\ref{sfig:13}B shows that at least they have the same sign.

We have also applied the models of Eqs.~(\ref{eq:cobb_douglas_scaling}) and (\ref{eq:translog_scaling}) to MSAs, $\mu$SAs, and CBSAs data. Supplementary Figure~\ref{sfig:14} shows that these models do not represent an improved description when compared with Eq.~(\ref{eq:density_scaling}). For CBSAs and MSAs, Eq.~(\ref{eq:cobb_douglas_scaling}) has about the same predictive power as Eq.~(\ref{eq:density_scaling}). Supplementary Figure~\ref{sfig:15} compares the exponents $\beta_P$ and $\beta_A$ obtained from MSAs, $\mu$SAs and CBSAs with the CCA values for $l=5$~km. We notice that $\beta_P$ is not so different from $\beta$ for these functional city definitions, indicating that the confounding effect of the area is much weaker when compared with the CCA results. This fact reinforces the idea that MSAs and $\mu$SAs areas are not good predictors for CO$_2$ emissions. We also observe that $\beta_P$ is much larger for the functional definitions, and that $\beta_A$ is smaller than the values obtained from the CCA approach. In spite of all discrepancies, the results for MSAs, $\mu$SAs and CBSAs also indicate that population has more impact on emissions than density because $|\beta_P+\beta_A|>|\beta_A|$. On the other hand, $\beta_P+\beta_A\approx1$ for the functional definitions, while $\beta_P+\beta_A<1$ for the CCA approach. The approximate constant returns to scale observed for the functional city definitions is also likely to be related to disaggregation and aggregation effects that we previously discussed.

Finally, it is worth remarking that the CCA is also likely to suffer from disaggregation or aggregation effects, as is the case of administrative or functional city definitions. However, differently from such ad hoc definitions, CCA allows us to quantify the impact of such effects by changing the threshold distance $l$ and to verify that our conclusions are robust under different values of $l$.

\subsection*{Cobb-Douglas and the urban scaling models}\label{sec:connecting_cobb}
To relate the Cobb-Douglas model [Eq.~(\ref{eq:cobb_douglas_scaling})] with the urban scaling [Eq.~(\ref{eq:urban_scaling})] and the per capita density scaling [Eq.~(\ref{eq:density_scaling})], we first rewrite Eq.~(\ref{eq:urban_scaling}) as
\begin{equation}\label{eq:urban_scaling_bar}
C\sim P^{\beta} = P^{\beta_1} P^{\beta_2}\,, 
\end{equation}
where $\beta = \beta_1 + \beta_2$ stands for the same urban scaling exponent as in Eq.~(\ref{eq:urban_scaling}). Now, we replace the right-most $P$ in the previous equation by the allometry relation between population and area~\cite{Stewart179Suggested,BattyF2011}, $P \sim A^\delta$ (where $\delta$ is another power-law exponent), leading to
\begin{equation}
C\sim P^{\beta_1} A^{\beta_2 \delta}\,,
\end{equation}
which has the same form as Eq.~(\ref{eq:cobb_douglas_scaling}), that is, $\beta_1=\beta_P$ and $\beta_2 \delta = \beta_A$. Consequently,
\begin{equation}
\beta = \beta_P + \frac{\beta_A}{\delta}\,,
\end{equation}
and the two approaches are uniquely related only if there is another constraint for the exponents. One possibility is by imposing constant returns to scale in the Cobb-Douglas model ($\beta_A+\beta_P=1$), which leads to
\begin{equation}\label{eq:connection_cobb_doublas_and_scalings}
\beta_P=\frac{\beta \delta-1}{\delta-1} \quad \text{and} \quad \beta_A=\frac{\delta(\beta-1)}{1-\delta}\,,
\end{equation}
for $\delta\neq 1$; while for $\delta=1$, we obtain $\beta_P=1$ and $\beta_A=0$. Thus, the existence of an additional constraint for the parameters $\beta_P$ and $\beta_A$ implies that the Cobb-Douglas model is equivalent to Eqs.~(\ref{eq:urban_scaling}) and (\ref{eq:density_scaling}); otherwise, it represents a generalization.

The Cobb-Douglas model with $\beta_A+\beta_P=1$ is also related to scaling relationships between indicator density and population density~\cite{um2009scaling,HanleyLR2016,ribeiro2018unveiling}. To obtain this connection, we rewrite Eq.~(\ref{eq:cobb_douglas_scaling}) as
\begin{equation}
C \sim P^\theta A^{1-\theta}\,,
\end{equation}
where $\beta_P=\theta$ and $\beta_A = 1-\theta$. Next, we divide both sides by $A$
\begin{equation}\label{eq:CPtheta}
C/A \sim (P/A)^\theta\,,
\end{equation}
leading to a scaling relationship between CO$_2$ density and population density.

\subsection*{Analogy with the production functions}\label{sec:analogy}
As we have argued, our approach is inspired by the economic theory of production functions~\cite{heathfield1987introduction}. By following this analogy, we have considered the urban emissions as the output and population and area (or density) as the inputs of a two-factor production process mediated by cities. The mathematical formula that describes the possible relations between the inputs ($P$ and $A$) and the output ($C$) is the production function, that is, $C=F(P,A)$. The functional forms for $F$ used in our work comprise the most widely known and used production functions in economics and should be viewed as an empirical/phenomenological description (as it also happens in economics). In what follows, we summarize concepts from the economic theory of production functions that have been used in our work.

Elasticity of scale. The elasticity of scale $\varepsilon $ is the ratio between a proportionate change in the output (emissions) and a proportionate change in the inputs (population and area), that is, $\varepsilon = (dF/F)/(dP/P)$, where $dP/P=dA/A$ represents the proportionate change in the inputs. This measure quantifies the impact of changing population and area on emissions.

Technical rate of substitution. The technical rate of substitution measures the rate at which an input must change in response to a change in the other input so that the output remains constant. In absolute value, it represents the slope of the isoquants of the production function. For instance, assuming a particular value for the output $F(P,A)=C_0$, the technical rate of substitution between area and population is $\delta_P^A = dA/dP$.

Elasticity of substitution. The elasticity of substitution $\sigma$ somehow summarizes the shape of an isoquant. It is defined as the ratio between a proportionate change in the inputs and the associated proportionate change in the slope of the isoquant. Mathematically, we write $\sigma = \frac{[d(P/A)/(P/A)]}{[d(dP/dA)/(dP/dA)]}=\frac{d\log(P/A)}{d\log(dP/dA)}$, where the numerator represents the proportionate change in the inputs and the denominator the proportionate change in the slope of the isoquant. This measure quantifies the efficiency at which population and area substitute each other.

Marginal products. The marginal product of an input is defined as the (infinitesimal) change in the output resulting from a (infinitesimal) change in one of the inputs. For instance, the marginal product of population is $\Delta_P = dF/dP$ (it can also be defined in terms of logarithmic quantities: $\Delta_P = d\log F/d\log P$).

We summarize all these properties calculated for Cobb-Douglas [Eq.~(\ref{eq:cobb_douglas_scaling})] and translog [Eq.~(\ref{eq:translog_scaling})] models in Supplementary Table~\ref{stab:2}.

\subsection*{Fitting models with the ridge regression approach}\label{sec:fitting}
As we have discussed in the main text, multicollinearity is present in the models of Eqs.~(\ref{eq:cobb_douglas_scaling}) and (\ref{eq:translog_scaling}). This effect happens when at least two predictors in a multiple linear regression are correlated to each other~\cite{hastie_elements_2016,hoerl1970ridge}. Under this situation and depending on the degree of correlation among the predictors, ordinary least-squares estimates of the parameters can be unstable against minor changes in the input data and also display large standard errors. To better illustrate this problem, consider the simple linear model
\begin{equation}
y \sim a_1 x_1 + a_2 x_2\,,
\end{equation}
where $y$ is the response variable, $x_1$ and $x_2$ are the predictors, and $a_1$ and $a_2$ are the linear coefficients. The least-squares estimator for the parameters is usually written as $\bm{a}=\icol{a_1\\a_2} = (\bm{X}^T \bm{X})^{-1} \bm{X}^T \bm{y}$, where $\bm{y}=\icol{y_1^{(1)}\\\vdots\\y^{(n)}}$ is a $n\times1$ vector of the response variables, $\bm{X}=\icol{x_1^{(1)}~x_2^{(1)}\\\vdots\hspace{0.6cm}\vdots\\x_1^{(n)}~x_2^{(n)}}$ is a $n\times2$ matrix of the regressors, and $n$ is the number of observations. If the values of predictors are strongly correlated, the inversion of the matrix $\bm{X}^T \bm{X}$ can become unstable, and consequently lead to unstable estimates for the linear coefficients. 

To account for the multicollinearity problem, we have fitted Eqs.~(\ref{eq:cobb_douglas_scaling}) and (\ref{eq:translog_scaling}) by using the ridge regression approach~\cite{hastie_elements_2016,hoerl1970ridge}. This method solves the matrix inversion problem by adding a constant $\lambda$ to the diagonal elements of $\bm{X}^T \bm{X}$, so that the ridge estimator for the linear coefficients is $\bm{a}=(\bm{X}^T \bm{X} + \lambda \bm{I})^{-1} \bm{X}^T \bm{y}$, where $\bm{I}$ is the identity matrix. The ridge estimation is equivalent to finding the optimal linear coefficients that minimize the residual sum of squares plus a penalty term (also called regularization parameter) proportional to the sum of the squares of the linear coefficients~\cite{hastie_elements_2016,hoerl1970ridge}, that is, finding the $\bm{a}$ that minimizes the objective function $\lVert \bm{y} - \bm{X}\bm{a}\rVert^2+ \lambda \lVert \bm{a} \rVert^2$. The optimal value of $\lambda$ is usually unknown in practice and needs to be estimated from data. To do so, we have used the approach of searching for the value of $\lambda$ that minimizes the mean squared error (MSE) in a leave-one-out cross validation strategy. In this approach, we estimate $\bm{a}$ (for a given $\lambda$) using all data except for one point that is used for calculating the squared error. This process is repeated until every data point is used exactly once for estimating the squared error, and then we calculate the value of the MSE for a given $\lambda$. The optimal value of $\lambda=\lambda^{*}$ is the one that minimizes the average value of the MSE estimated with the leave-one-out cross validation method. We have also standardized all predictors before searching for the optimal value $\lambda^{*}$. This is a common practice when dealing with regularization methods and ensures that the penalty term is uniformly applied to the predictors, that is, the normalization makes the scale of the predictors comparable and prevents variables with distinct ranges from having uneven penalization. 

The standardized version of Eq.~(\ref{eq:cobb_douglas_scaling}) can be written as
\begin{equation}\label{eq:cobb_douglas_scaling_std}
\log C \sim \tilde{\beta}_P \, \widehat{\log P} + \tilde{\beta}_A\, \widehat{\log A}\,,
\end{equation}
where
\begin{equation}\label{eq:normalization}
\widehat{\log P} = \frac{\log P - \mu_P}{\sigma_P}~~\text{and}~~
\widehat{\log A} = \frac{\log A - \mu_A}{\sigma_A}\,.
\end{equation}
In addition, $\mu_P$ is the average value of $\log P$ ($\mu_P=\frac{1}{n}\sum \log P$), $\sigma_P$ is the standard deviation  of $\log P$ [$\sigma^2_P=\frac{1}{n-1}\sum (\log P - \mu_P)^2$], $\mu_A$ is the average value of $\log A$ ($\mu_A=\frac{1}{n}\sum \log A$), and $\sigma_A$ is the standard deviation  of $\log A$ [$\sigma^2_A=\frac{1}{n-1}\sum (\log A - \mu_A)^2$]. It is worth remarking that the values of $\widehat{\log P}$ and $\widehat{\log A}$ are invariant against changes in the scale of $P$ and $A$, that is, their values do not change under the transformations $P\to \nu P$ and $A\to \nu A$, where $\nu$ is a positive constant. The same invariance holds for $\sigma_P$ and $\sigma_A$, whereas the average values change according to $\mu_P \to \log \nu + \mu_P$ and $\mu_A \to \log \nu + \mu_A$.

The connection between the parameters of the standardized model ($\tilde{\beta}_P$ and $\tilde{\beta}_A$) and the usual ones ($\beta_P$ and $\beta_A$) is obtained by plugging Eq.~(\ref{eq:normalization}) into Eq.~(\ref{eq:cobb_douglas_scaling_std}), collecting the terms multiplying $\log P$ and $\log A$, and then directly comparing the results with Eq.~(\ref{eq:cobb_douglas_scaling}). By following this approach, we find that
\begin{equation}\label{eq:connection_cobb_douglas}
\beta_P = \frac{\tilde{\beta}_P}{\sigma_P\, \sigma_A}~~\text{and}~~\beta_A = \frac{\tilde{\beta}_A}{\sigma_P\, \sigma_A}\,,
\end{equation}
where we observe that $\beta_P$ and $\beta_A$ are independent of the units of $P$ and $A$ (as they should be, since $\sigma_P$ and $\sigma_A$ are scale invariants). Supplementary Figure~\ref{sfig:16} illustrates how the ridge regression approach is applied to the model of Eq.~(\ref{eq:cobb_douglas_scaling_std}). Supplementary Figure~\ref{sfig:16}A shows the dependence of the MSE on the values of $\lambda$, from which we obtain $\lambda^{*}=9.78$. Supplementary Figure~\ref{sfig:16}B shows the dependence of the parameters $\tilde{\beta}_A$ and $\tilde{\beta}_P$, whose values for the optimal regularization term ($\lambda=\lambda^{*}$) are $\tilde{\beta}_P=0.37\pm0.02$ and $\tilde{\beta}_A=0.24\pm0.02$, which correspond to ${\beta}_P=0.31\pm0.01$ and ${\beta}_A=0.45\pm0.03$. The errors in $\tilde{\beta}_P$ and $\tilde{\beta}_A$ stand for the standard deviation of their values estimated over 1000 random samples with replacement of the data, as shown in Supplementary Figure~\ref{sfig:17}. The errors in ${\beta}_P$ and ${\beta}_A$ are calculated with common error propagation formulas. Moreover, the $p$-values of permutation tests reject the null hypothesis that these parameters are equal to zero.

In the case of Eq.~(\ref{eq:translog_scaling}), its standardized version can be rewritten as
\begin{equation}\label{eq:translog_scaling_std}
\log{C} \sim \tilde{\beta}_P \,\widehat{\log P} + \tilde{\beta}_A\, \widehat{\log A} + \tilde{\beta}_C\, \widehat{\log (P)}\; \widehat{\log (A)}\,,
\end{equation}
where the connecting formulas
\begin{equation}\label{eq:connection_translog}
\beta_P = \frac{\sigma_A \tilde{\beta}_P - \mu_A \tilde{\beta}_C}{\sigma_P\, \sigma_A}\,,~~\beta_A = \frac{\sigma_P \tilde{\beta}_A - \mu_P \tilde{\beta}_C}{\sigma_P\, \sigma_A}\,,~~\text{and}~~\beta_C = \frac{\tilde{\beta}_C}{\sigma_P\, \sigma_A}\,,
\end{equation}
are obtained as in the previous case. Unlike the models of Eqs.~(\ref{eq:urban_scaling}), (\ref{eq:density_scaling}), and (\ref{eq:cobb_douglas_scaling}), the generalization expressed by Eq.~(\ref{eq:translog_scaling}) is not scale-invariant and thus its parameters depend on the measurement units. In particular, if the area is rescaled by a factor $\nu$ ($A\to \nu A$), $\beta_P$ is incremented by the factor $-\beta_C \log\nu$ and $\beta_A$ remains unchanged. Similarly, if population is rescaled by $\nu$ ($P\to \nu P$), $\beta_A$ is also incremented by $-\beta_C \log\nu$ and $\beta_P$ remains unchanged. Only $\beta_C$ is invariant against scale changes in $P$ and $A$. Thus, all interpretations related to the behavior of the emissions obtained from the model Eq.~(\ref{eq:translog_scaling}) involve the assumption that area is expressed in units of km$^2$ and raw population counts. Supplementary Figure~\ref{sfig:18} illustrates how the ridge regression approach is applied to the model of Eq.~(\ref{eq:translog_scaling_std}). Supplementary Figure~\ref{sfig:18}A shows the dependence of the MSE on $\lambda$, from which $\lambda^{*}=8.67$ is obtained. Supplementary Figure~\ref{sfig:18}B shows the behavior of the model parameters as a function of $\lambda$. This approach yields $\tilde{\beta}_P=0.40\pm0.02$, $\tilde{\beta}_A=0.17\pm0.02$, and $\tilde{\beta}_C=0.044\pm0.006$ for $\lambda=\lambda^{*}$, which correspond to $\beta_P=0.28\pm 0.02$, $\beta_A = 0.14\pm0.05$, and $\beta_C=0.07\pm0.01$. The standard errors are calculated as in the previous case and the $p$-values of the permutation tests reject the null hypothesis that the model parameters are equal to zero (Supplementary Figure~\ref{sfig:3}).

In addition to the models of Eqs.~(\ref{eq:cobb_douglas_scaling}) and (\ref{eq:translog_scaling}), we have further tested for a more general translog model (full translog model) having quadratic terms in $\log P$ and $\log A$, that is,
\begin{equation}\label{eq:translog_scaling_full}
\begin{split}
\log{C} & \sim \beta_P \log P + \beta_A \log A + \beta_C \log (P) \log (A)\\
& + \beta_P' (\log P)^2 + \beta_A' (\log A)^2\,,
\end{split}
\end{equation}
where $\beta_P'$ and $\beta_A'$ are additional parameters. This expression can also be related to the CES production by applying the Taylor series expansion to Eq.~(\ref{eq:CES_scaling}) around the point $\gamma=0$ (the Kmenta approximation~\cite{kmenta1967estimation}). We have fitted Eq.~(\ref{eq:translog_scaling_full}) by following the same procedure used for Eqs.~(\ref{eq:cobb_douglas_scaling}) and (\ref{eq:translog_scaling}). In particular, Supplementary Figure~\ref{sfig:19} illustrates how the ridge approach is applied to this model and Supplementary Figure~\ref{sfig:20} shows the best fitting parameters and their estimated errors. However, as shown in Supplementary Figure~\ref{sfig:2}, the full translog model of Eq.~(\ref{eq:translog_scaling_full}) does not improve the goodness of the fit when compared with Eq.~(\ref{eq:translog_scaling}). 

\section*{Data availability}
The dataset used in this study were obtained from Gudipudi et al.~\cite{Gudipudi2016}, which in turn rely on freely available data obtained from the Global Rural-Urban Mapping Project (GRUMP)~\cite{GRUMP}, the Global Land Cover Dataset (GLC)~\cite{GLC}, and the Vulcan Project~\cite{gurney2010vulcan}. All data supporting the findings of this study are available from the corresponding authors on reasonable request.

\section*{Code availability}
The code used for analyzing data is available from the corresponding authors on reasonable request.

\section*{Author contributions}
H.V.R, D.R., J.P.K. designed research, performed research, analyzed data, and wrote the paper.

\section*{Competing interests}
The authors declare no competing interests.

\begin{acknowledgments}
We thank R. Gudipudi for providing the data and useful discussions.
This work emerged from ideas discussed at the symposium \emph{Cities as Complex Systems} (Hanover, July 13th-15th, 2016) which was generously funded by Volkswagen Foundation. H.V.~Ribeiro thanks the financial support of CNPq (grants 303250/2015-1, 407690/2018-2, and 303121/2018-1). D.~Rybski is grateful to the Leibniz Association (project IMPETUS) for financially supporting this research. J.P.~Kropp acknowledges financial cooperation support (Brazil: KoUP1 -- Remus) by University of Potsdam.
\end{acknowledgments}

\linespread{1.01}
\bibliographystyle{naturemag}
\bibliography{CO2_pop_area.bib}

\begin{thebibliography}{10}
\expandafter\ifx\csname url\endcsname\relax
  \def\url#1{\texttt{#1}}\fi
\expandafter\ifx\csname urlprefix\endcsname\relax\def\urlprefix{URL }\fi
\providecommand{\bibinfo}[2]{#2}
\providecommand{\eprint}[2][]{\url{#2}}

\bibitem{Schellnhuber2006}
\bibinfo{author}{Schellnhuber, H.~J.}, \bibinfo{author}{Cramer, W.},
  \bibinfo{author}{Nakicenovic, N.}, \bibinfo{author}{Wigley, T.} \&
  \bibinfo{author}{Yohe, G.}
\newblock \emph{\bibinfo{title}{Avoiding Dangerous Climate Change}}
  (\bibinfo{publisher}{Cambridge University Press}, \bibinfo{address}{New
  York}, \bibinfo{year}{2006}).

\bibitem{EschHHKRZDS2017}
\bibinfo{author}{Esch, T.} \emph{et~al.}
\newblock \bibinfo{title}{Breaking new ground in mapping human settlements from
  space -- the global urban footprint}.
\newblock \emph{\bibinfo{journal}{ISPRS Journal of Photogrammetry and Remote
  Sensing}} \textbf{\bibinfo{volume}{134}}, \bibinfo{pages}{30--42}
  (\bibinfo{year}{2017}).

\bibitem{johansson2012global}
\bibinfo{author}{Johansson, T.~B.}, \bibinfo{author}{Patwardhan, A.~P.},
  \bibinfo{author}{Naki{\'c}enovi{\'c}, N.} \&
  \bibinfo{author}{Gomez-Echeverri, L.}
\newblock \emph{\bibinfo{title}{Global Energy Assessment: {Toward} a
  Sustainable Future}} (\bibinfo{publisher}{Cambridge University Press},
  \bibinfo{address}{Cambridge}, \bibinfo{year}{2012}).

\bibitem{seto2014human}
\bibinfo{author}{Seto, K.} \emph{et~al.}
\newblock \bibinfo{title}{{Chapter 12 - Human Settlements, Infrastructure and
  Spatial Planning}}.
\newblock In \emph{\bibinfo{booktitle}{Climate Change 2014: {Mitigation} of
  Climate Change. IPCC Working Group III Contribution to AR5}}
  (\bibinfo{publisher}{Cambridge University Press}, \bibinfo{address}{New
  York}, \bibinfo{year}{2014}).

\bibitem{NewmanK1989}
\bibinfo{author}{Newman, P. W.~C.} \& \bibinfo{author}{Kenworthy, J.~R.}
\newblock \bibinfo{title}{Gasoline consumption and cities}.
\newblock \emph{\bibinfo{journal}{Journal of the American Planning
  Association}} \textbf{\bibinfo{volume}{55}}, \bibinfo{pages}{24--37}
  (\bibinfo{year}{1989}).

\bibitem{lariviere1999modelling}
\bibinfo{author}{Lariviere, I.} \& \bibinfo{author}{Lafrance, G.}
\newblock \bibinfo{title}{Modelling the electricity consumption of cities:
  {Effect} of urban density}.
\newblock \emph{\bibinfo{journal}{Energy Economics}}
  \textbf{\bibinfo{volume}{21}}, \bibinfo{pages}{53--66}
  (\bibinfo{year}{1999}).

\bibitem{brown2009geography}
\bibinfo{author}{Brown, M.~A.}, \bibinfo{author}{Southworth, F.} \&
  \bibinfo{author}{Sarzynski, A.}
\newblock \bibinfo{title}{The geography of metropolitan carbon footprints}.
\newblock \emph{\bibinfo{journal}{Policy and Society}}
  \textbf{\bibinfo{volume}{27}}, \bibinfo{pages}{285--304}
  (\bibinfo{year}{2009}).

\bibitem{FragkiasLSS2013}
\bibinfo{author}{Fragkias, M.}, \bibinfo{author}{Lobo, J.},
  \bibinfo{author}{Strumsky, D.} \& \bibinfo{author}{Seto, K.~C.}
\newblock \bibinfo{title}{Does size matter? {S}caling of {CO2} emissions and
  {U.S.} urban areas}.
\newblock \emph{\bibinfo{journal}{PLOS ONE}} \textbf{\bibinfo{volume}{8}},
  \bibinfo{pages}{e64727} (\bibinfo{year}{2013}).

\bibitem{OliveiraAM2014}
\bibinfo{author}{Oliveira, E.~A.}, \bibinfo{author}{Andrade, J.~S.} \&
  \bibinfo{author}{Makse, H.~A.}
\newblock \bibinfo{title}{Large cities are less green}.
\newblock \emph{\bibinfo{journal}{Scientific Reports}}
  \textbf{\bibinfo{volume}{4}}, \bibinfo{pages}{4235} (\bibinfo{year}{2014}).

\bibitem{tamayao2014us}
\bibinfo{author}{Tamayao, M.}, \bibinfo{author}{Blackhurst, M.} \&
  \bibinfo{author}{Matthews, H.}
\newblock \bibinfo{title}{Do {US} metropolitan core counties have lower scope 1
  and 2 {CO2} emissions than less urbanized counties?}
\newblock \emph{\bibinfo{journal}{Environmental Research Letters}}
  \textbf{\bibinfo{volume}{9}}, \bibinfo{pages}{104011} (\bibinfo{year}{2014}).

\bibitem{LoufB2014SciRep}
\bibinfo{author}{Louf, R.} \& \bibinfo{author}{Barthelemy, M.}
\newblock \bibinfo{title}{How congestion shapes cities: {From} mobility
  patterns to scaling}.
\newblock \emph{\bibinfo{journal}{Scientific Reports}}
  \textbf{\bibinfo{volume}{4}}, \bibinfo{pages}{5561} (\bibinfo{year}{2014}).

\bibitem{jones2014spatial}
\bibinfo{author}{Jones, C.} \& \bibinfo{author}{Kammen, D.~M.}
\newblock \bibinfo{title}{Spatial distribution of {US} household carbon
  footprints reveals suburbanization undermines greenhouse gas benefits of
  urban population density}.
\newblock \emph{\bibinfo{journal}{Environmental Science \& Technology}}
  \textbf{\bibinfo{volume}{48}}, \bibinfo{pages}{895--902}
  (\bibinfo{year}{2014}).

\bibitem{mohajeri2015co}
\bibinfo{author}{Mohajeri, N.}, \bibinfo{author}{Gudmundsson, A.} \&
  \bibinfo{author}{French, J.~R.}
\newblock \bibinfo{title}{{CO2} emissions in relation to street-network
  configuration and city size}.
\newblock \emph{\bibinfo{journal}{Transportation Research Part D: Transport and
  Environment}} \textbf{\bibinfo{volume}{35}}, \bibinfo{pages}{116--129}
  (\bibinfo{year}{2015}).

\bibitem{ye2015sustainable}
\bibinfo{author}{Ye, H.} \emph{et~al.}
\newblock \bibinfo{title}{A sustainable urban form: {The} challenges of
  compactness from the viewpoint of energy consumption and carbon emission}.
\newblock \emph{\bibinfo{journal}{Energy and Buildings}}
  \textbf{\bibinfo{volume}{93}}, \bibinfo{pages}{90--98}
  (\bibinfo{year}{2015}).

\bibitem{Gudipudi2016}
\bibinfo{author}{Gudipudi, R.}, \bibinfo{author}{Fluschnik, T.},
  \bibinfo{author}{Ros, A. G.~C.}, \bibinfo{author}{Walther, C.} \&
  \bibinfo{author}{Kropp, J.~P.}
\newblock \bibinfo{title}{City density and {CO2} efficiency}.
\newblock \emph{\bibinfo{journal}{Energy Policy}}
  \textbf{\bibinfo{volume}{91}}, \bibinfo{pages}{352--361}
  (\bibinfo{year}{2016}).

\bibitem{chang2017there}
\bibinfo{author}{Chang, Y.~S.}, \bibinfo{author}{Lee, Y.~J.} \&
  \bibinfo{author}{Choi, S. S.~B.}
\newblock \bibinfo{title}{Is there more traffic congestion in larger cities? -
  {Scaling} analysis of the 101 largest {US} urban centers}.
\newblock \emph{\bibinfo{journal}{Transport Policy}}
  \textbf{\bibinfo{volume}{59}}, \bibinfo{pages}{54--63}
  (\bibinfo{year}{2017}).

\bibitem{rybski2017cities}
\bibinfo{author}{Rybski, D.} \emph{et~al.}
\newblock \bibinfo{title}{Cities as nuclei of sustainability?}
\newblock \emph{\bibinfo{journal}{Environment and Planning B}}
  \textbf{\bibinfo{volume}{44}}, \bibinfo{pages}{425--440}
  (\bibinfo{year}{2017}).

\bibitem{chen2017coupling}
\bibinfo{author}{Chen, S.} \& \bibinfo{author}{Chen, B.}
\newblock \bibinfo{title}{Coupling of carbon and energy flows in cities: {A}
  meta-analysis and nexus modelling}.
\newblock \emph{\bibinfo{journal}{Applied Energy}}
  \textbf{\bibinfo{volume}{194}}, \bibinfo{pages}{774--783}
  (\bibinfo{year}{2017}).

\bibitem{muller2017does}
\bibinfo{author}{Muller, N.~Z.} \& \bibinfo{author}{Jha, A.}
\newblock \bibinfo{title}{Does environmental policy affect scaling laws between
  population and pollution? {Evidence} from {American} metropolitan areas}.
\newblock \emph{\bibinfo{journal}{PLOS ONE}} \textbf{\bibinfo{volume}{12}},
  \bibinfo{pages}{e0181407} (\bibinfo{year}{2017}).

\bibitem{gudipudi2019efficient}
\bibinfo{author}{Gudipudi, R.} \emph{et~al.}
\newblock \bibinfo{title}{The efficient, the intensive, and the productive:
  {Insights} from urban {Kaya} scaling}.
\newblock \emph{\bibinfo{journal}{Applied Energy}}
  \textbf{\bibinfo{volume}{236}}, \bibinfo{pages}{155--162}
  (\bibinfo{year}{2019}).

\bibitem{bettencourt2007growth}
\bibinfo{author}{Bettencourt, L. M.~A.}, \bibinfo{author}{Lobo, J.},
  \bibinfo{author}{Helbing, D.}, \bibinfo{author}{K{\"u}hnert, C.} \&
  \bibinfo{author}{West, G.~B.}
\newblock \bibinfo{title}{Growth, innovation, scaling, and the pace of life in
  cities}.
\newblock \emph{\bibinfo{journal}{Proceedings of the National Academy of
  Sciences}} \textbf{\bibinfo{volume}{104}}, \bibinfo{pages}{7301--7306}
  (\bibinfo{year}{2007}).

\bibitem{bettencourt2010unified}
\bibinfo{author}{Bettencourt, L. M.~A.} \& \bibinfo{author}{West, G.}
\newblock \bibinfo{title}{A unified theory of urban living}.
\newblock \emph{\bibinfo{journal}{Nature}} \textbf{\bibinfo{volume}{467}},
  \bibinfo{pages}{912--913} (\bibinfo{year}{2010}).

\bibitem{baur2013urban}
\bibinfo{author}{Baur, A.~H.}, \bibinfo{author}{Thess, M.},
  \bibinfo{author}{Kleinschmit, B.} \& \bibinfo{author}{Creutzig, F.}
\newblock \bibinfo{title}{Urban climate change mitigation in europe: looking at
  and beyond the role of population density}.
\newblock \emph{\bibinfo{journal}{Journal of Urban Planning and Development}}
  \textbf{\bibinfo{volume}{140}}, \bibinfo{pages}{04013003}
  (\bibinfo{year}{2013}).

\bibitem{Stewart179Suggested}
\bibinfo{author}{Stewart, J.~Q.}
\newblock \bibinfo{title}{Suggested principles of `social physics'}.
\newblock \emph{\bibinfo{journal}{Science}} \textbf{\bibinfo{volume}{106}},
  \bibinfo{pages}{179--180} (\bibinfo{year}{1947}).

\bibitem{BattyF2011}
\bibinfo{author}{Batty, M.} \& \bibinfo{author}{Ferguson, P.}
\newblock \bibinfo{title}{Defining city size}.
\newblock \emph{\bibinfo{journal}{Environment and Planning B}}
  \textbf{\bibinfo{volume}{38}}, \bibinfo{pages}{753--756}
  (\bibinfo{year}{2011}).

\bibitem{heathfield1987introduction}
\bibinfo{author}{David F.~Heathfield, S.~W.}
\newblock \emph{\bibinfo{title}{An Introduction to Cost and Production
  Functions}} (\bibinfo{publisher}{Macmillan}, \bibinfo{address}{London},
  \bibinfo{year}{1987}).

\bibitem{rozenfeld2008laws}
\bibinfo{author}{Rozenfeld, H.~D.} \emph{et~al.}
\newblock \bibinfo{title}{Laws of population growth}.
\newblock \emph{\bibinfo{journal}{Proceedings of the National Academy of
  Sciences}} \textbf{\bibinfo{volume}{105}}, \bibinfo{pages}{18702--18707}
  (\bibinfo{year}{2008}).

\bibitem{CobbD1928}
\bibinfo{author}{Cobb, C.~W.} \& \bibinfo{author}{Douglas, P.~H.}
\newblock \bibinfo{title}{A theory of production}.
\newblock \emph{\bibinfo{journal}{American Economic Review}}
  \textbf{\bibinfo{volume}{18}}, \bibinfo{pages}{139--165}
  (\bibinfo{year}{1928}).

\bibitem{hoerl1970ridge}
\bibinfo{author}{Hoerl, A.~E.} \& \bibinfo{author}{Kennard, R.~W.}
\newblock \bibinfo{title}{Ridge regression: {Biased} estimation for
  nonorthogonal problems}.
\newblock \emph{\bibinfo{journal}{Technometrics}}
  \textbf{\bibinfo{volume}{12}}, \bibinfo{pages}{55--67}
  (\bibinfo{year}{1970}).

\bibitem{hastie_elements_2016}
\bibinfo{author}{Hastie, T.}, \bibinfo{author}{Tibshirani, R.} \&
  \bibinfo{author}{Friedman, J.}
\newblock \emph{\bibinfo{title}{The {Elements} of {Statistical} {Learning}:
  {Data} {Mining}, {Inference}, and {Prediction}}}
  (\bibinfo{publisher}{Springer}, \bibinfo{address}{New York},
  \bibinfo{year}{2016}), \bibinfo{edition}{2nd} edn.

\bibitem{arrow1961capital}
\bibinfo{author}{Arrow, K.~J.}, \bibinfo{author}{Chenery, H.~B.},
  \bibinfo{author}{Minhas, B.~S.} \& \bibinfo{author}{Solow, R.~M.}
\newblock \bibinfo{title}{Capital-labor substitution and economic efficiency}.
\newblock \emph{\bibinfo{journal}{The Review of Economics and Statistics}}
  \textbf{\bibinfo{volume}{43}}, \bibinfo{pages}{225--250}
  (\bibinfo{year}{1961}).

\bibitem{christensen1973transcendental}
\bibinfo{author}{Christensen, L.~R.}, \bibinfo{author}{Jorgenson, D.~W.} \&
  \bibinfo{author}{Lau, L.~J.}
\newblock \bibinfo{title}{Transcendental logarithmic production frontiers}.
\newblock \emph{\bibinfo{journal}{The Review of Economics and Statistics}}
  \textbf{\bibinfo{volume}{55}}, \bibinfo{pages}{28--45}
  (\bibinfo{year}{1973}).

\bibitem{dodman2011forces}
\bibinfo{author}{Dodman, D.}
\newblock \bibinfo{title}{Forces driving urban greenhouse gas emissions}.
\newblock \emph{\bibinfo{journal}{Current Opinion in Environmental
  Sustainability}} \textbf{\bibinfo{volume}{3}}, \bibinfo{pages}{121--125}
  (\bibinfo{year}{2011}).

\bibitem{dietz1997effects}
\bibinfo{author}{Dietz, T.} \& \bibinfo{author}{Rosa, E.~A.}
\newblock \bibinfo{title}{Effects of population and affluence on {CO2}
  emissions}.
\newblock \emph{\bibinfo{journal}{Proceedings of the National Academy of
  Sciences}} \textbf{\bibinfo{volume}{94}}, \bibinfo{pages}{175--179}
  (\bibinfo{year}{1997}).

\bibitem{Chertow_IPAT_2000}
\bibinfo{author}{Chertow, M.~R.}
\newblock \bibinfo{title}{The {IPAT} equation and its variants}.
\newblock \emph{\bibinfo{journal}{Journal of Industrial Ecology}}
  \textbf{\bibinfo{volume}{4}}, \bibinfo{pages}{13--29} (\bibinfo{year}{2000}).

\bibitem{Waggoner2002framework}
\bibinfo{author}{Waggoner, P.~E.} \& \bibinfo{author}{Ausubel, J.~H.}
\newblock \bibinfo{title}{A framework for sustainability science: {A} renovated
  {IPAT} identity}.
\newblock \emph{\bibinfo{journal}{Proceedings of the National Academy of
  Sciences}} \textbf{\bibinfo{volume}{99}}, \bibinfo{pages}{7860--7865}
  (\bibinfo{year}{2002}).

\bibitem{ArcauteHFYJB2014}
\bibinfo{author}{Arcaute, E.} \emph{et~al.}
\newblock \bibinfo{title}{Constructing cities, deconstructing scaling laws}.
\newblock \emph{\bibinfo{journal}{Journal of the Royal Society Interface}}
  \textbf{\bibinfo{volume}{12}}, \bibinfo{pages}{20140745}
  (\bibinfo{year}{2014}).

\bibitem{cai2014urban}
\bibinfo{author}{Cai, B.} \& \bibinfo{author}{Zhang, L.}
\newblock \bibinfo{title}{Urban {CO2} emissions in {China}: {Spatial} boundary
  and performance comparison}.
\newblock \emph{\bibinfo{journal}{Energy Policy}}
  \textbf{\bibinfo{volume}{66}}, \bibinfo{pages}{557--567}
  (\bibinfo{year}{2014}).

\bibitem{wang2017urban}
\bibinfo{author}{Wang, Y.}, \bibinfo{author}{Yang, L.}, \bibinfo{author}{Han,
  S.}, \bibinfo{author}{Li, C.} \& \bibinfo{author}{Ramachandra, T.}
\newblock \bibinfo{title}{Urban {CO2} emissions in {Xi’an} and {Bangalore} by
  commuters: {Implications} for controlling urban transportation carbon dioxide
  emissions in developing countries}.
\newblock \emph{\bibinfo{journal}{Mitigation and Adaptation Strategies for
  Global Change}} \textbf{\bibinfo{volume}{22}}, \bibinfo{pages}{993--1019}
  (\bibinfo{year}{2017}).

\bibitem{wang2014changing}
\bibinfo{author}{Wang, Y.}, \bibinfo{author}{Hayashi, Y.},
  \bibinfo{author}{Chen, J.} \& \bibinfo{author}{Li, Q.}
\newblock \bibinfo{title}{Changing urban form and transport {CO2} emissions:
  {An} empirical analysis of {Beijing}, {China}}.
\newblock \emph{\bibinfo{journal}{Sustainability}}
  \textbf{\bibinfo{volume}{6}}, \bibinfo{pages}{4558--4579}
  (\bibinfo{year}{2014}).

\bibitem{yang2019urban}
\bibinfo{author}{Yang, L.}, \bibinfo{author}{Wang, Y.}, \bibinfo{author}{Han,
  S.} \& \bibinfo{author}{Liu, Y.}
\newblock \bibinfo{title}{Urban transport carbon dioxide ({CO2}) emissions by
  commuters in rapidly developing cities: {The} comparative study of {Beijing}
  and {Xi’an} in {China}}.
\newblock \emph{\bibinfo{journal}{Transportation Research Part D: Transport and
  Environment}} \textbf{\bibinfo{volume}{68}}, \bibinfo{pages}{65--83}
  (\bibinfo{year}{2019}).

\bibitem{GRUMP}
\bibinfo{title}{{Center for International Earth Science Information Network -
  CIESIN - Columbia University, International Food Policy Research Institute -
  IFPRI, The World Bank, and Centro Internacional de Agricultura Tropical -
  CIAT. 2011. Global Rural-Urban Mapping Project, Version 1 (GRUMPv1):
  {Population} Density Grid. Palisades, NY: NASA Socioeconomic Data and
  Applications Center (SEDAC).}}
\newblock \bibinfo{howpublished}{{Available:}
  \url{http://sedac.ciesin.columbia.edu/data/set/grump-v1-population-density}}.
\newblock \bibinfo{note}{{Accessed:} 14 Mar 2019}.

\bibitem{GLC}
\bibinfo{title}{{Global Land Cover 2000 Project (GLC 2000)}}.
\newblock \bibinfo{howpublished}{{Available:}
  \url{http://forobs.jrc.ec.europa.eu/products/glc2000/glc2000.php}}.
\newblock \bibinfo{note}{{Accessed:} 14 Mar 2019}.

\bibitem{gurney2010vulcan}
\bibinfo{author}{Gurney, K.~R.} \emph{et~al.}
\newblock \bibinfo{title}{Vulcan science methods documentation, version 2.0}.
\newblock \bibinfo{howpublished}{{Available:}
  \url{http://vulcan.project.asu.edu}}.
\newblock \bibinfo{note}{{Accessed:} 14 Mar 2019}.

\bibitem{bettencourt2013hypothesis}
\bibinfo{author}{Bettencourt, L.}, \bibinfo{author}{Lobo, J.} \&
  \bibinfo{author}{Youn, H.}
\newblock \bibinfo{title}{The hypothesis of urban scaling: formalization,
  implications and challenges}.
\newblock \emph{\bibinfo{journal}{arXiv:1301.5919}}  (\bibinfo{year}{2013}).
\newblock \urlprefix\url{https://arxiv.org/abs/1301.5919}.
\newblock \bibinfo{note}{{Accessed:} 14 Mar 2019}.

\bibitem{um2009scaling}
\bibinfo{author}{Um, J.}, \bibinfo{author}{Son, S.-W.}, \bibinfo{author}{Lee,
  S.-I.}, \bibinfo{author}{Jeong, H.} \& \bibinfo{author}{Kim, B.~J.}
\newblock \bibinfo{title}{Scaling laws between population and facility
  densities}.
\newblock \emph{\bibinfo{journal}{Proceedings of the National Academy of
  Sciences}} \textbf{\bibinfo{volume}{106}}, \bibinfo{pages}{14236--14240}
  (\bibinfo{year}{2009}).

\bibitem{HanleyLR2016}
\bibinfo{author}{Hanley, Q.~S.}, \bibinfo{author}{Lewis, D.} \&
  \bibinfo{author}{Ribeiro, H.~V.}
\newblock \bibinfo{title}{Rural to urban population density scaling of crime
  and property transactions in {English} and {Welsh} {Parliamentary}
  {Constituencies}}.
\newblock \emph{\bibinfo{journal}{PLOS ONE}} \textbf{\bibinfo{volume}{11}},
  \bibinfo{pages}{e0149546} (\bibinfo{year}{2016}).

\bibitem{ribeiro2018unveiling}
\bibinfo{author}{Ribeiro, H.~V.}, \bibinfo{author}{Hanley, Q.~S.} \&
  \bibinfo{author}{Lewis, D.}
\newblock \bibinfo{title}{Unveiling relationships between crime and property in
  {England} and {Wales} via density scale-adjusted metrics and network tools}.
\newblock \emph{\bibinfo{journal}{PLOS ONE}} \textbf{\bibinfo{volume}{13}},
  \bibinfo{pages}{e0192931} (\bibinfo{year}{2018}).

\bibitem{kmenta1967estimation}
\bibinfo{author}{Kmenta, J.}
\newblock \bibinfo{title}{On estimation of the {CES} production function}.
\newblock \emph{\bibinfo{journal}{International Economic Review}}
  \textbf{\bibinfo{volume}{8}}, \bibinfo{pages}{180--189}
  (\bibinfo{year}{1967}).

\end{thebibliography}
\clearpage

\setcounter{page}{1}
\setcounter{figure}{0}
\makeatletter
\renewcommand{\figurename}{Supplementary Figure}
\renewcommand{\tablename}{Supplementary Table}
\renewcommand{\thetable}{\@arabic\c@table}
\clearpage

\onecolumngrid
\begin{center}
\large{Supplementary Information for}\\
\vskip1pc
\large{\bf Effects of changing population or density on urban carbon dioxide emissions}\\
\vskip1pc
\normalsize{Haroldo V.\ Ribeiro et al., Nature Communications, 2019.}
\end{center}
\clearpage

\begin{figure*}[!ht]
\centering
\includegraphics[width=1\linewidth]{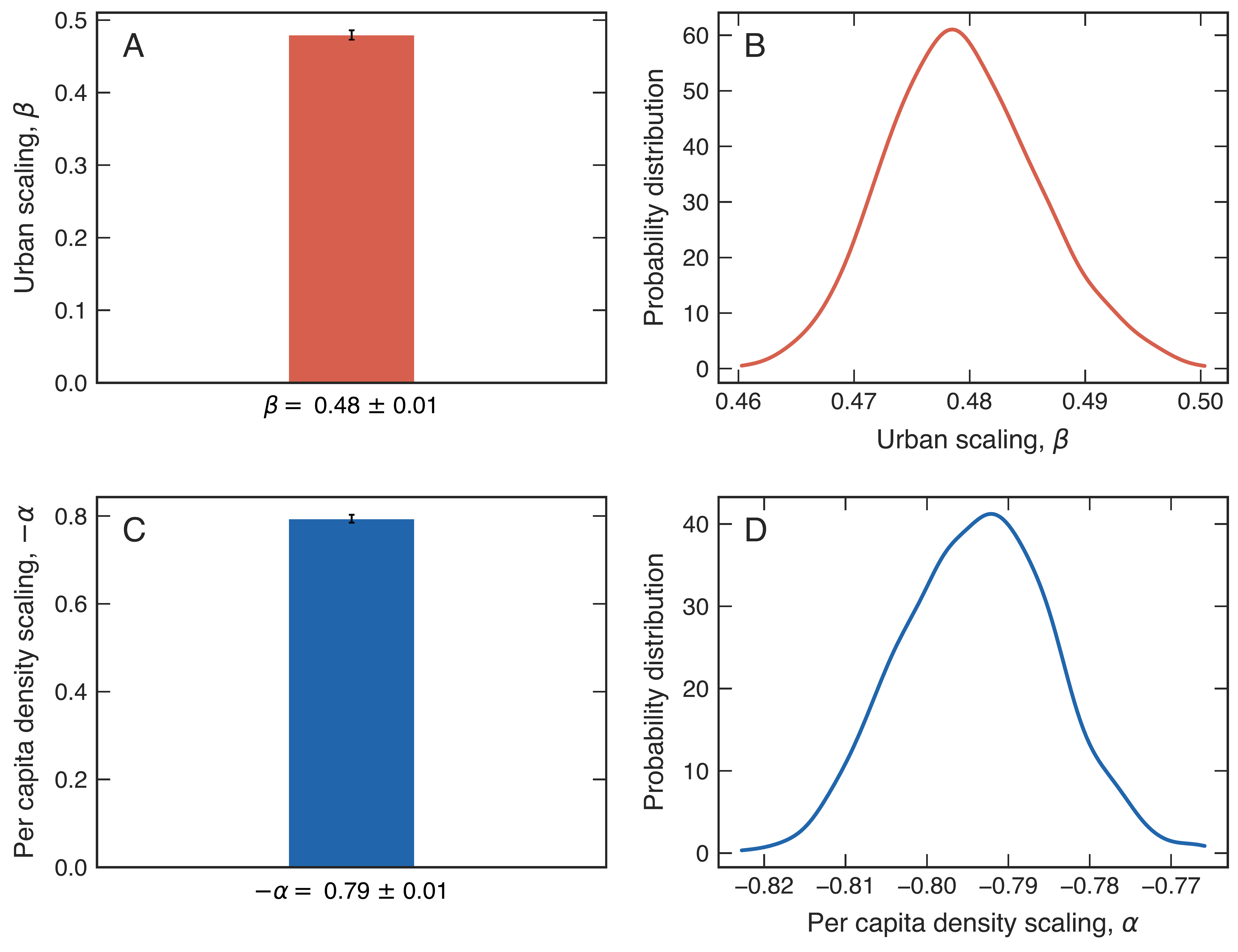}
\caption{{\bf Estimating the errors in the parameters of Eqs.~(\ref{eq:urban_scaling}) [urban scaling] and (\ref{eq:density_scaling}) [per capita density scaling].} Bar plots in panels (A) and (C) show the average values of the parameters $\beta$ and $\alpha$ estimated after fitting the models to 1000 random samples (with replacement) of our data. In these panels, error bars stand for the standard deviation. Panels (B) and (D) show the probability distribution of the values of $\beta$ and $\alpha$ over all random samples. In both cases, the $p$-values of the permutation test are virtually zero, rejecting the null hypothesis that $\beta$ and $\alpha$ are zero.}
\label{sfig:1}
\end{figure*}

\begin{figure*}[!ht]
\centering
\includegraphics[width=1\linewidth]{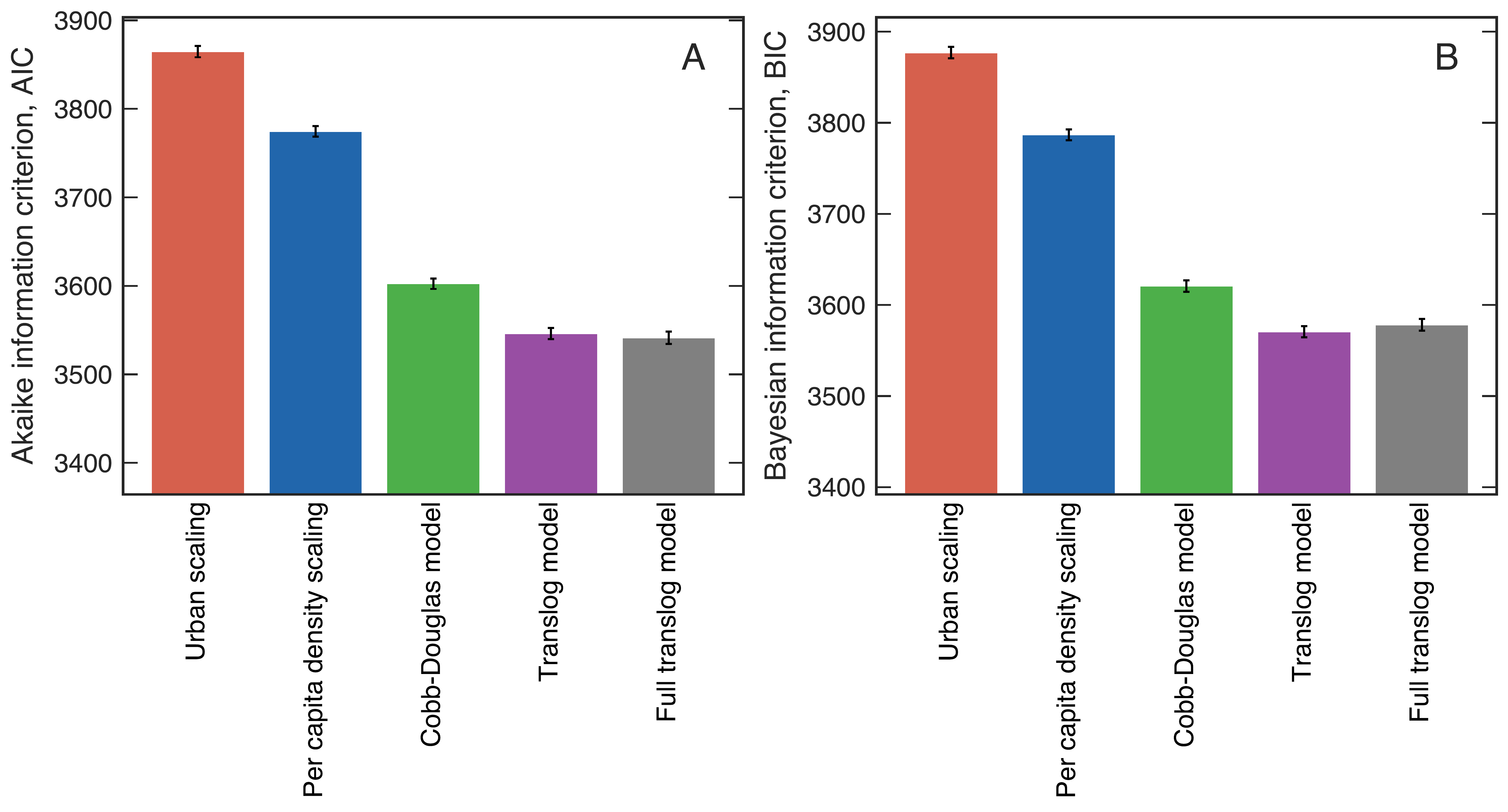}
\caption{{\bf Comparing the goodness of the fit among the models of Eqs.~(\ref{eq:urban_scaling}) [urban scaling], (\ref{eq:density_scaling}) [per capita density scaling], (\ref{eq:cobb_douglas_scaling}) [Cobb-Douglas model], (\ref{eq:translog_scaling}) [translog model], and (\ref{eq:translog_scaling_full}) [full translog model].} (A) Average Akaike information criterion (AIC) and (B) Bayesian information criterion (BIC) values for each model. These values were estimated from 1000 different random samplings (with replacement) of our data. The smaller the values of the AIC and BIC coefficients, the better is the quality of fit. Error bars are 95\% bootstrap confidence intervals. We note that the models of Eqs.~(\ref{eq:translog_scaling}) and (\ref{eq:translog_scaling_full}) provide significantly lower values when compared with all other models; however, no significant difference is observed between these two models. It is worth noting that these coefficients include a penalty term that depends on the number of parameters, allowing a fair comparison among models with different number of parameters. We further tested whether different linear regression fitting approaches improve the AIC and BIC for the urban scaling and per capita density scaling. We find that robust ordinary least squares, total least squares, and least absolute deviations regression do not improve the quality of fits of Eqs.~(\ref{eq:urban_scaling}) and (\ref{eq:density_scaling}).} 
\label{sfig:2}
\end{figure*}

\begin{figure*}[!ht]
\centering
\includegraphics[width=1\linewidth]{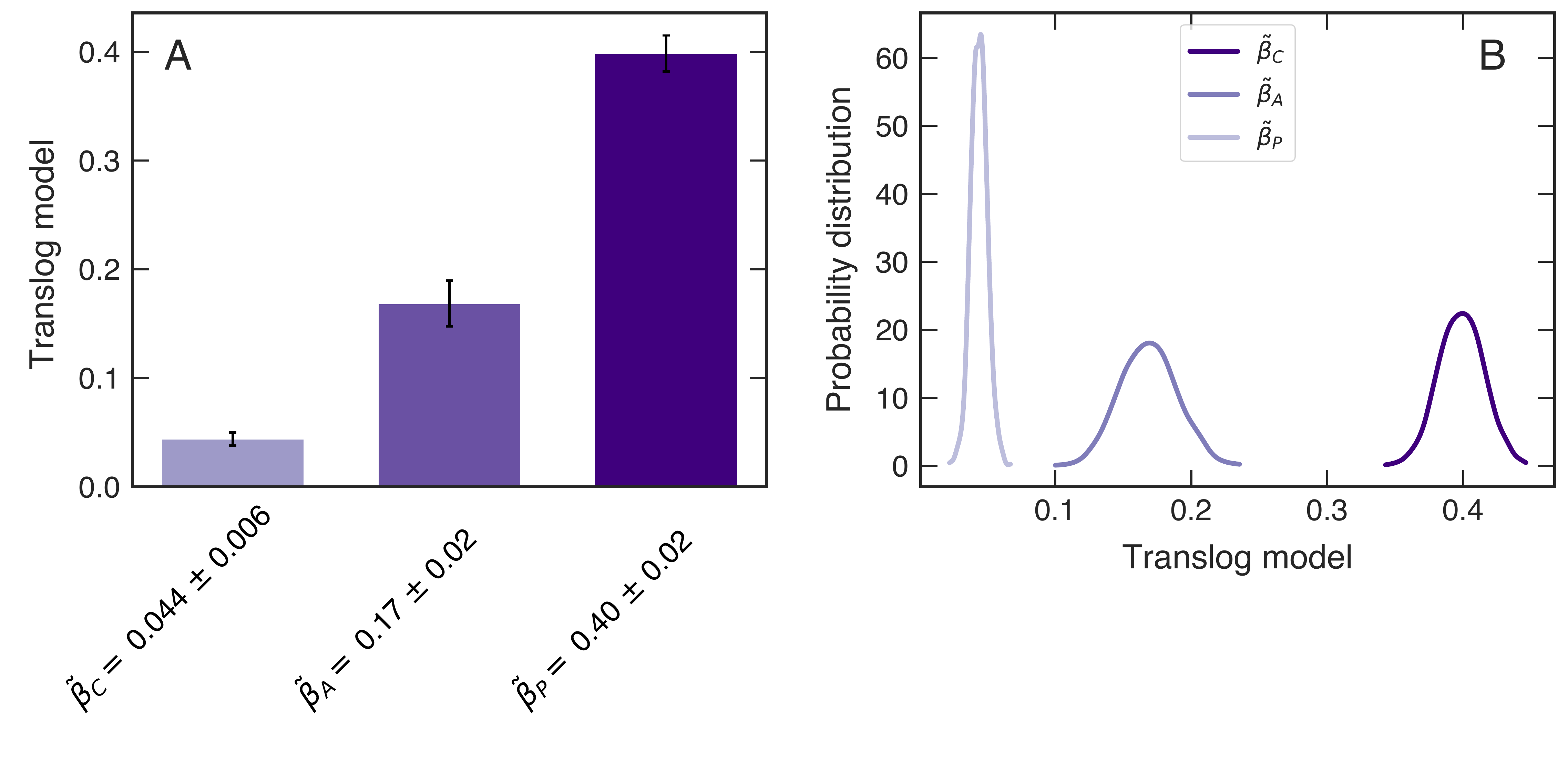}
\caption{{\bf Estimating the errors in the parameters of Eq.~(\ref{eq:translog_scaling}) [translog model].} (A) The bar plot shows the average values of the parameters $\tilde{\beta}_P$, $\tilde{\beta}_A$, and $\tilde{\beta}_C$ estimated after fitting the model to 1000 random samples (with replacement) of our data. Error bars stand for the standard deviation of these values. (B) Probability distribution of the values of $\tilde{\beta}_P$, $\tilde{\beta}_A$, and $\tilde{\beta}_C$ over all random samples. The permutation test on the model coefficients rejects the null hypothesis that they are equal to zero at the 95\% confidence level.
}
\label{sfig:3}
\end{figure*}

\begin{figure*}[!ht]
\centering
\includegraphics[width=1\linewidth]{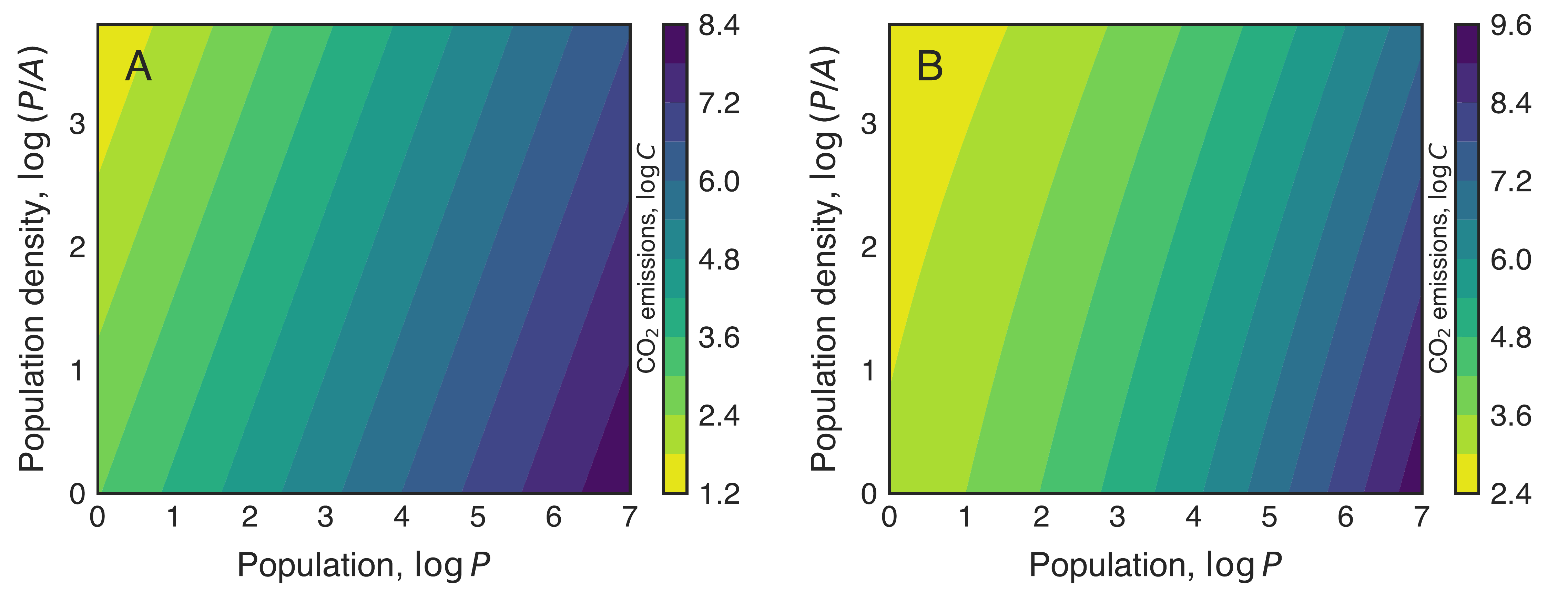}
\caption{{\bf Contour plots of Eqs.~(\ref{eq:cobb_douglas_scaling}) [Cobb-Douglas model] and (\ref{eq:translog_scaling}) [translog model] in terms of population density.} (A) Contour plot of Eq.~(\ref{eq:cobb_douglas_scaling}) rewritten in terms of population density, that is, $C \sim P^{\beta_P+\beta_A}\,(P/A)^{-\beta_A}$, with $\beta_P=0.31\pm0.01$ and $\beta_A=0.45\pm0.02$. (B) Contour plot of Eq.~(\ref{eq:translog_scaling}) rewritten in terms of population density, that is, $C\sim P^{\beta_P + \beta_A + \beta_C\log P}(P/A)^{-\beta_A-\beta_C\log P}$, with $\beta_P=0.28\pm0.02$, $\beta_A=0.14\pm0.05$, and $\beta_C=0.07\pm0.01$. We have employed base-$10$ logarithmic quantities in all panels.}
\label{sfig:4}
\end{figure*}

\begin{figure*}[!ht]
\centering
\includegraphics[width=1\linewidth]{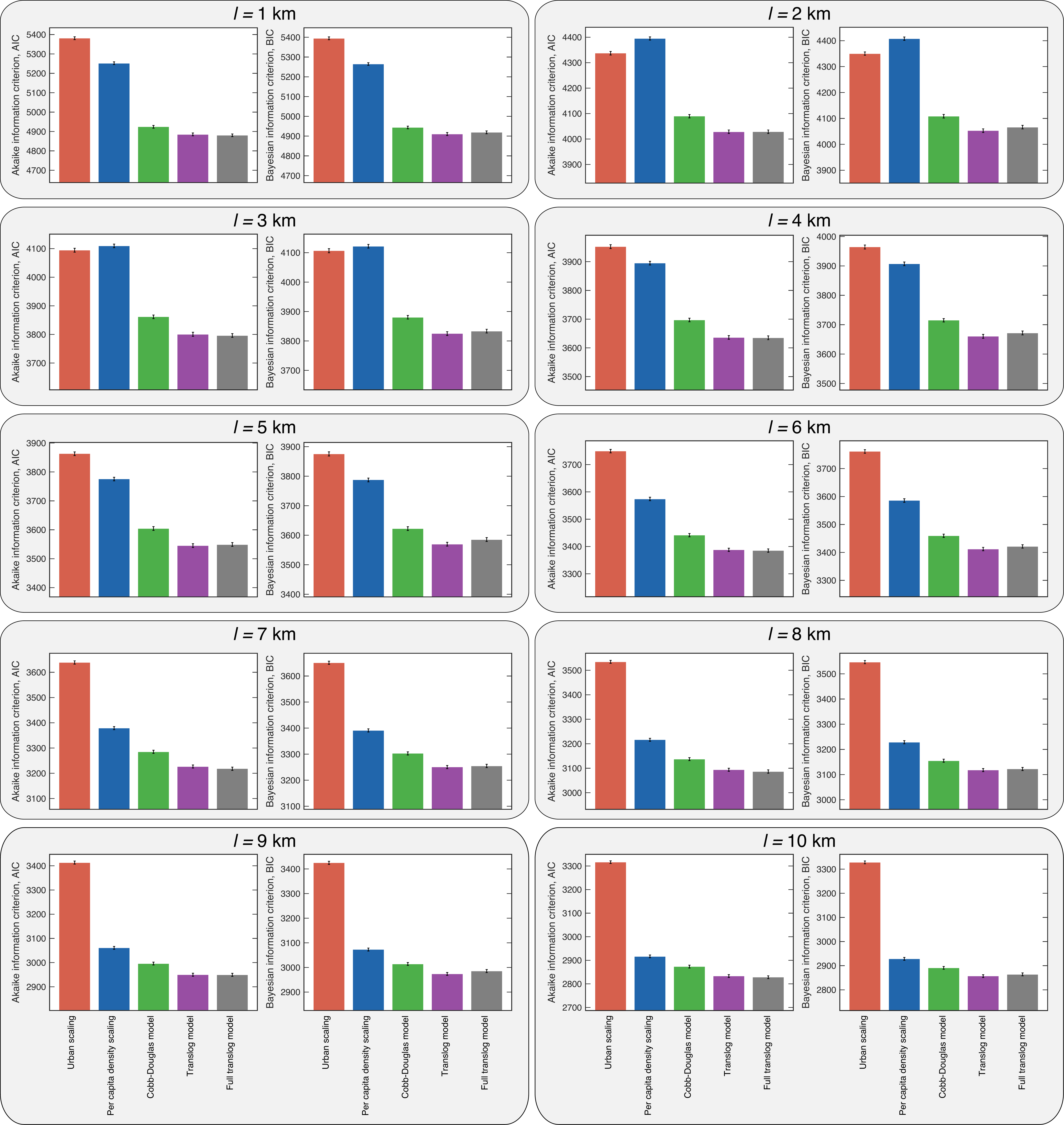}
\caption{{\bf Comparing the goodness of the fit among the models of Eqs.~(\ref{eq:urban_scaling}) [urban scaling], (\ref{eq:density_scaling}) [per capita density scaling], (\ref{eq:cobb_douglas_scaling}) [Cobb-Douglas model], (\ref{eq:translog_scaling}) [translog model], and (\ref{eq:translog_scaling_full}) [full translog model] under different values of the CCA threshold distance $l$.} Each panel shows the average Akaike information criterion (AIC) and the Bayesian information criterion (BIC) estimated from 1000 different random samplings (with replacement) of our data for different values of $l\in(1,2,3,4,5,6,7,8,9,10)$~km. In all plots, the error bars are 95\% bootstrap confidence intervals.  We note that the translog model [Eq.~(\ref{eq:translog_scaling})] always provides the best description regardless the value of $l$.}
\label{sfig:5}
\end{figure*}

\begin{figure*}[!ht]
\centering
\includegraphics[width=1\linewidth]{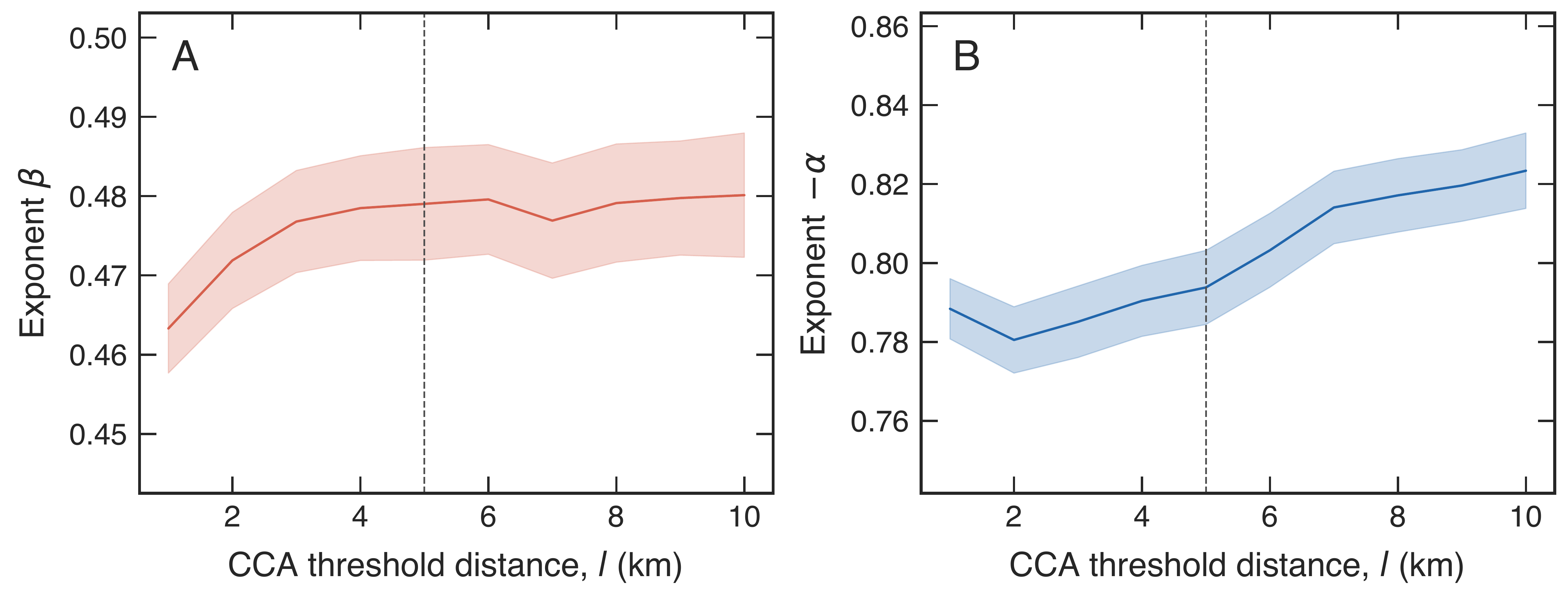}
\caption{{\bf The urban scaling exponents $\beta$ [Eq.~(\ref{eq:urban_scaling}), (A)] and $\alpha$ [Eq.~(\ref{eq:density_scaling}), (B)] as functions of the CCA threshold distance $l$.} In both plots, the shaded regions stand for the standard deviation in the parameters estimated after fitting the models to 1000 random samples (with replacement). The vertical lines indicate the values for $l=5$~km.
}
\label{sfig:6}
\end{figure*}

\begin{figure*}[!ht]
\centering
\includegraphics[width=1\linewidth]{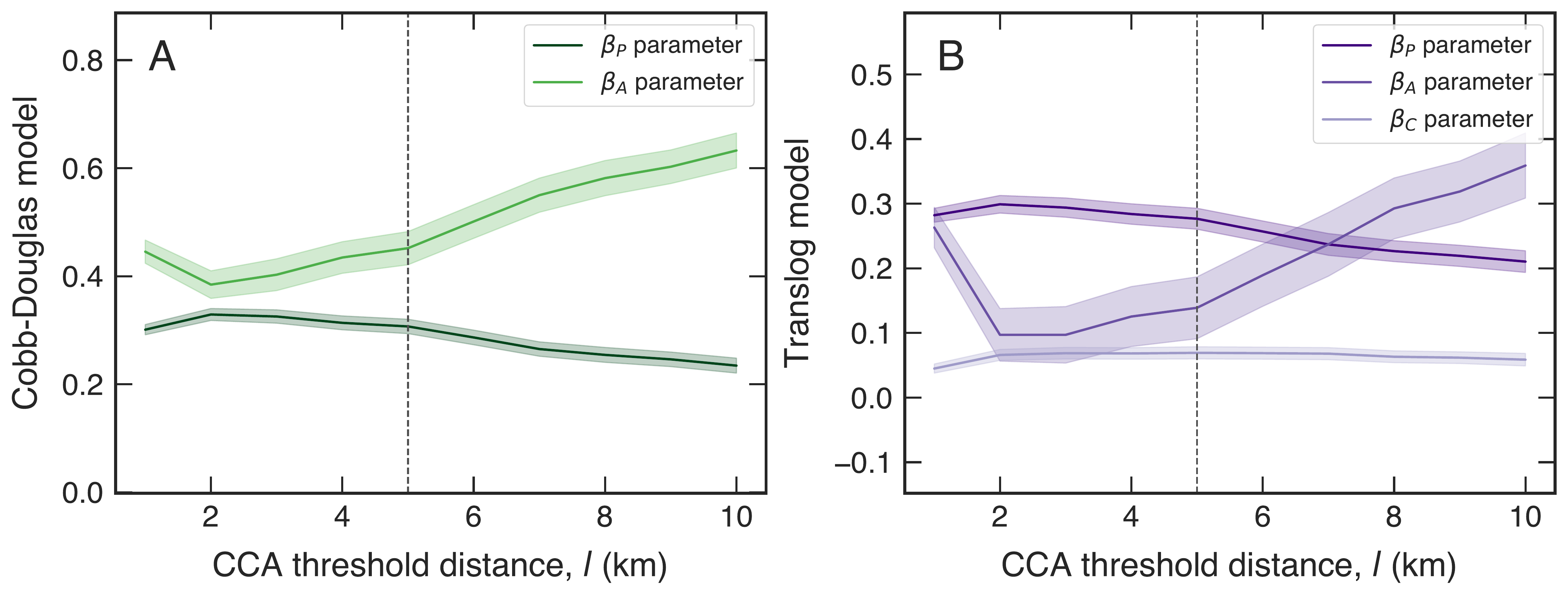}
\caption{{\bf The parameters of Cobb-Douglas [$\beta_P$ and $\beta_A$ in Eq.~(\ref{eq:cobb_douglas_scaling}), (A)] and translog [$\beta_P$, $\beta_A$, and $\beta_C$ in Eq.~(\ref{eq:translog_scaling}), (B)] models as functions of the CCA threshold distance $l$.} In both plots, the shaded regions stand for the standard deviation in the parameters estimated after fitting the models to 1000 random samples (with replacement). The vertical lines indicate the values for $l=5$~km.
}
\label{sfig:7}
\end{figure*}

\begin{figure*}[!ht]
\centering
\includegraphics[width=1\linewidth]{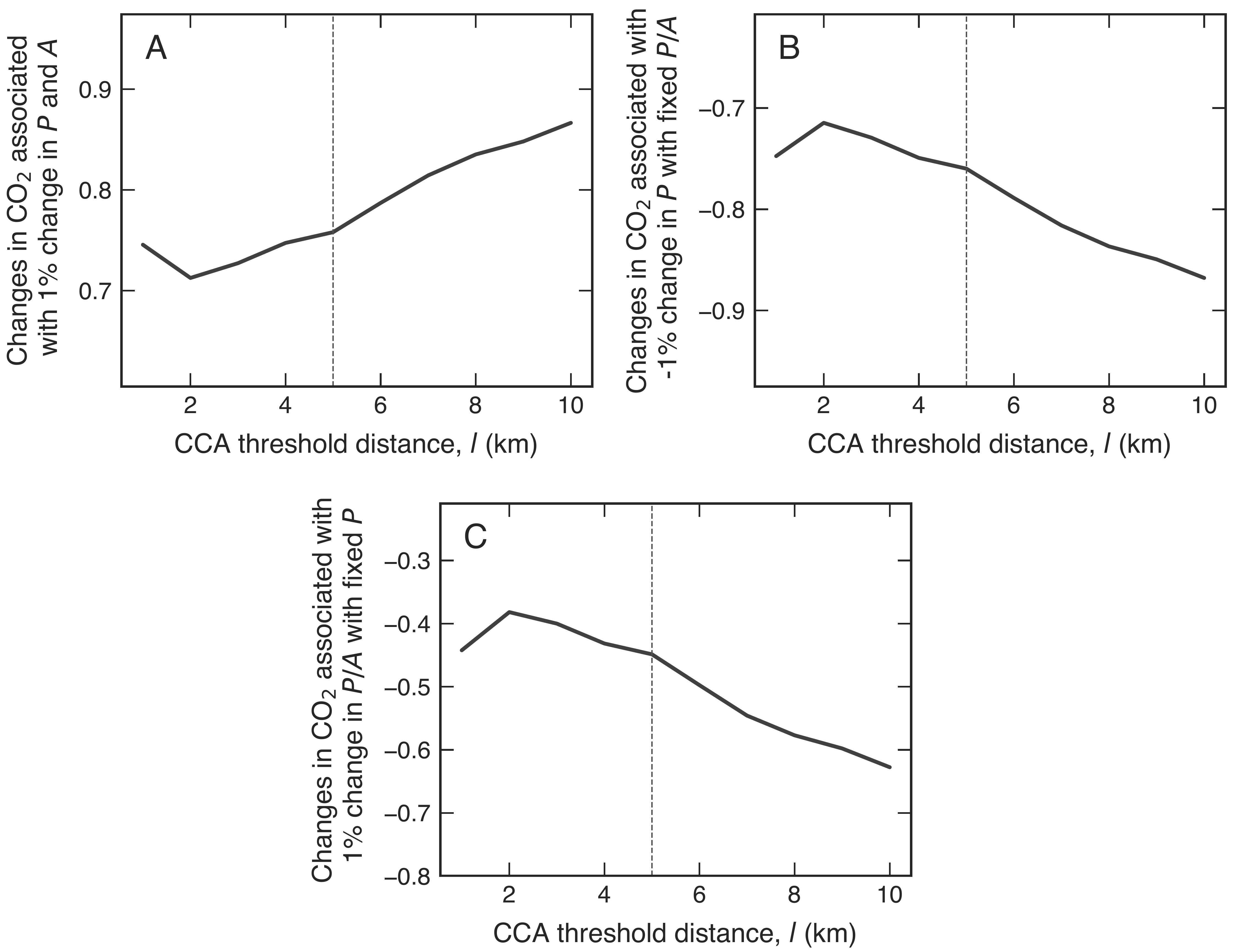}
\caption{{\bf Changes in the point estimates of the Cobb-Douglas model [Eq.~(\ref{eq:cobb_douglas_scaling})] reported in the manuscript against variations in CCA threshold distance $l$.} (A) The effect of a 1\% change in population and area on the emissions as a function of $l$. (B) The effect of a -1\% change in population (with fixed density) on the emissions as a function of $l$. (C) The effect of a 1\% change in population density (with fixed population) on the emissions as a function of $l$. The vertical lines indicate the values for $l=5$~km.
}
\label{sfig:8}
\end{figure*}

\begin{figure*}[!ht]
\centering
\includegraphics[width=1\linewidth]{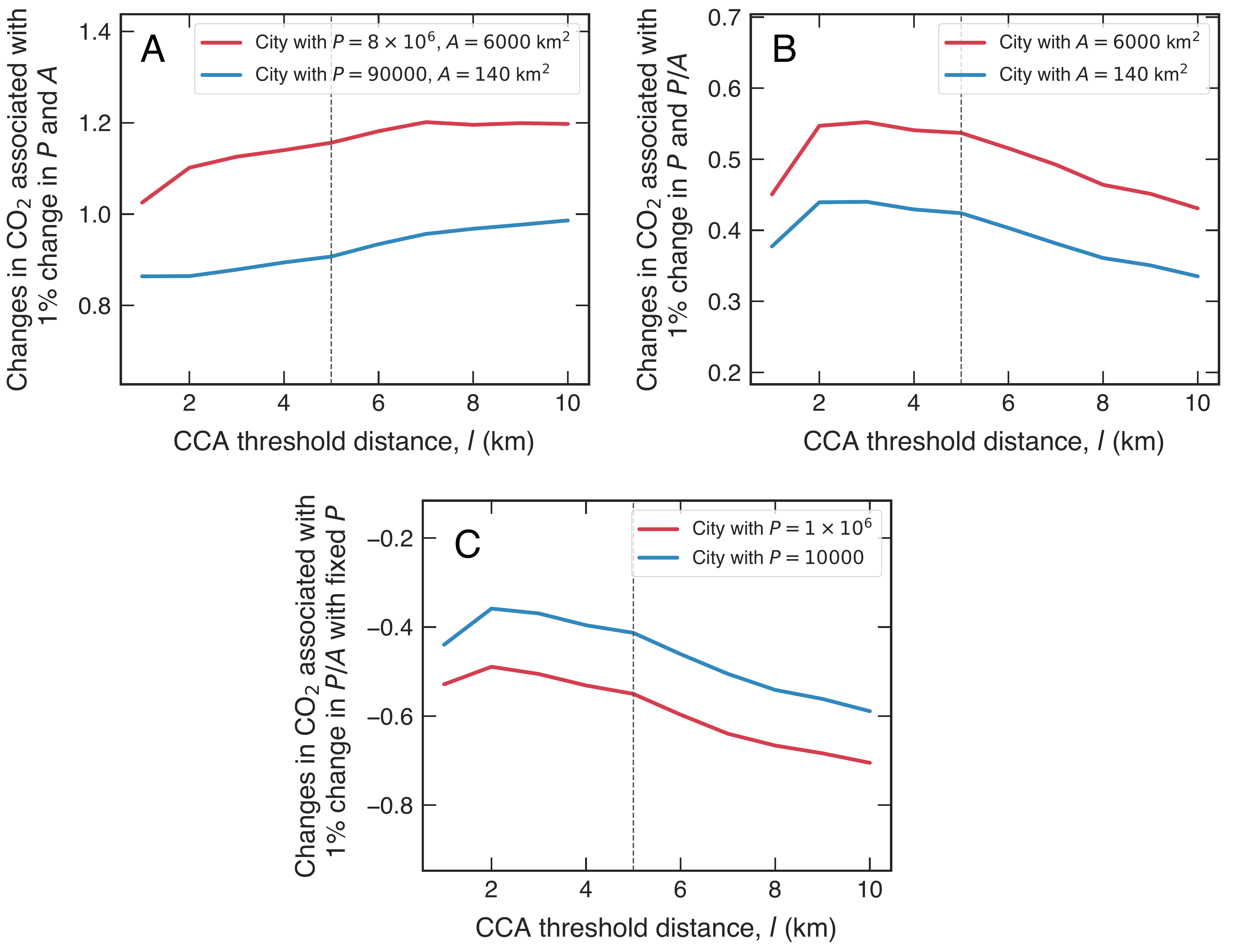}
\caption{{\bf Changes in the point estimates of the translog model [Eq.~(\ref{eq:translog_scaling})] reported in the manuscript against variations in the CCA threshold distance $l$.} (A) The effect of a 1\% change in population and area of two hypothetical cities with different initial values for $P$ and $A$ (as indicated within the plot) as function of $l$. (B) The effect of a 1\% change in population (with fixed density) of two hypothetical cities with different initial values for $A$ (as indicated within the plot) as function of $l$. (C) The effect of a 1\% change in population density (with fixed population) of two hypothetical cities with different initial values for $P$ (as indicated within the plot) as function of $l$. The vertical lines indicate the values for $l=5$~km.
}
\label{sfig:9}
\end{figure*}

\begin{figure*}[!ht]
\centering
\includegraphics[width=1\linewidth]{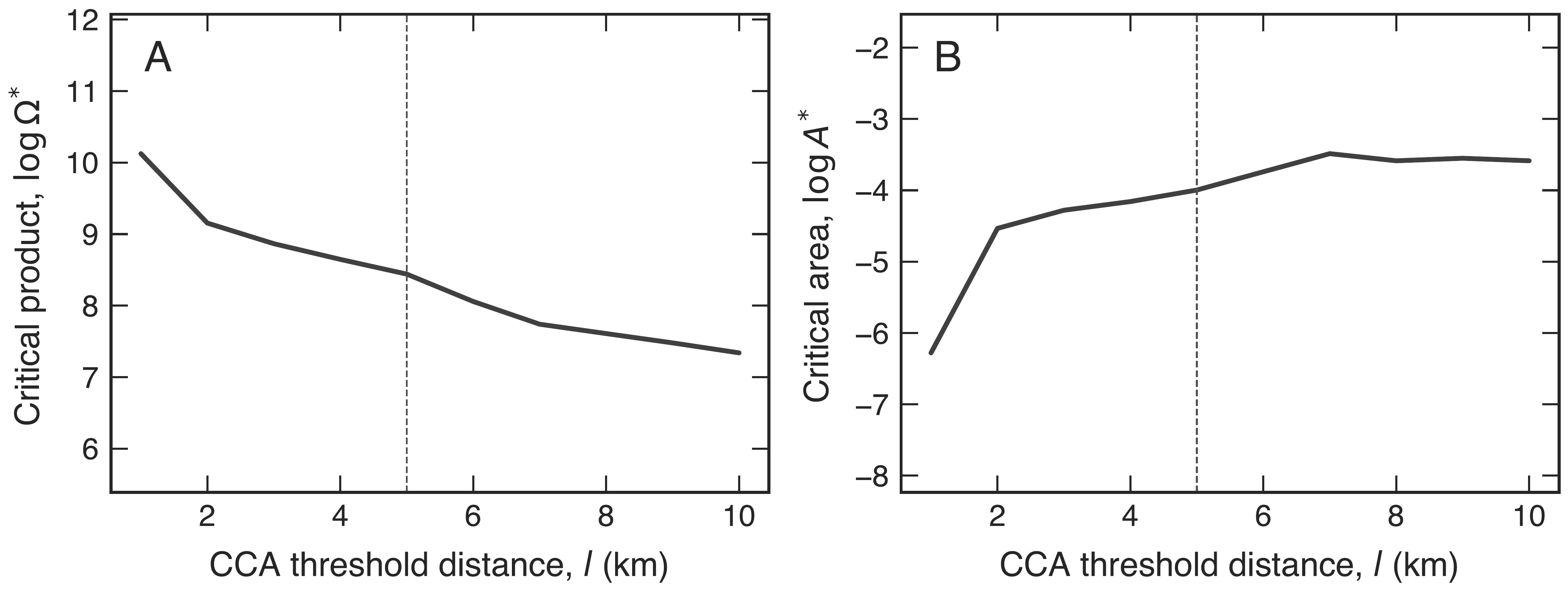}
\caption{{\bf Robustness of the critical values $\Omega^*$ and $A^*$ against variations in the CCA threshold distance $l$.} (A) The critical product $\Omega^*=10^{(1-\beta_P-\beta_A)/\beta_C}$  and (B) the critical area $A^*=10^{-\beta_P/\beta_C}$ as a function of $l$. The vertical lines indicate the values for $l=5$~km.
}
\label{sfig:10}
\end{figure*}

\begin{figure*}[!ht]
\centering
\includegraphics[width=1\linewidth]{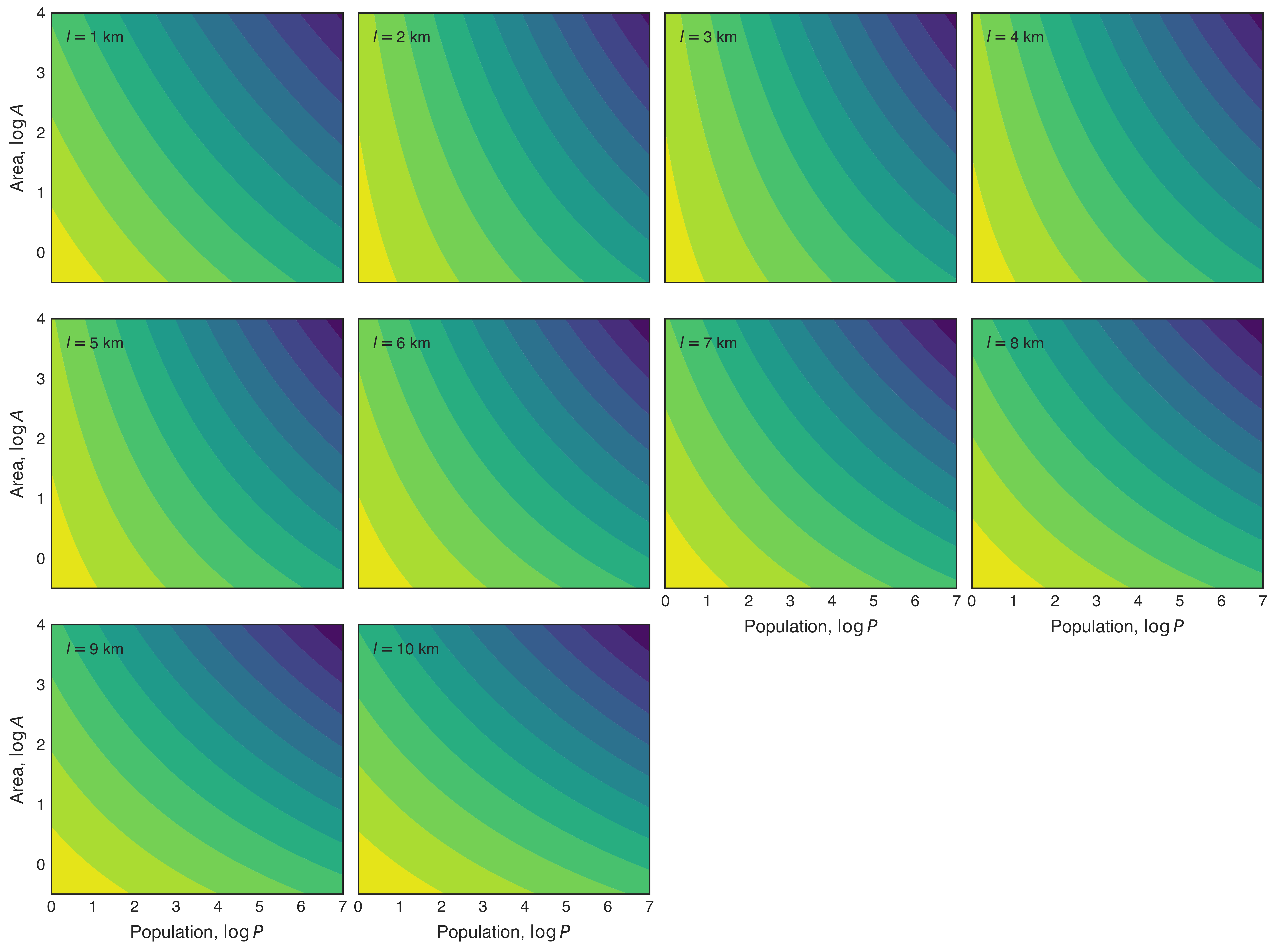}
\caption{{\bf The shape of the isoquants of the translog model Eq.~(\ref{eq:translog_scaling})] against variations in the CCA threshold distance $l$.} Each panel shows the contour plot of the translog model [Eq.~(\ref{eq:translog_scaling})] with the best fitting parameters for different values of $l$ (indicated in the plots).
}
\label{sfig:11}
\end{figure*}


\begin{figure*}[!ht]
\centering
\includegraphics[width=0.9\linewidth]{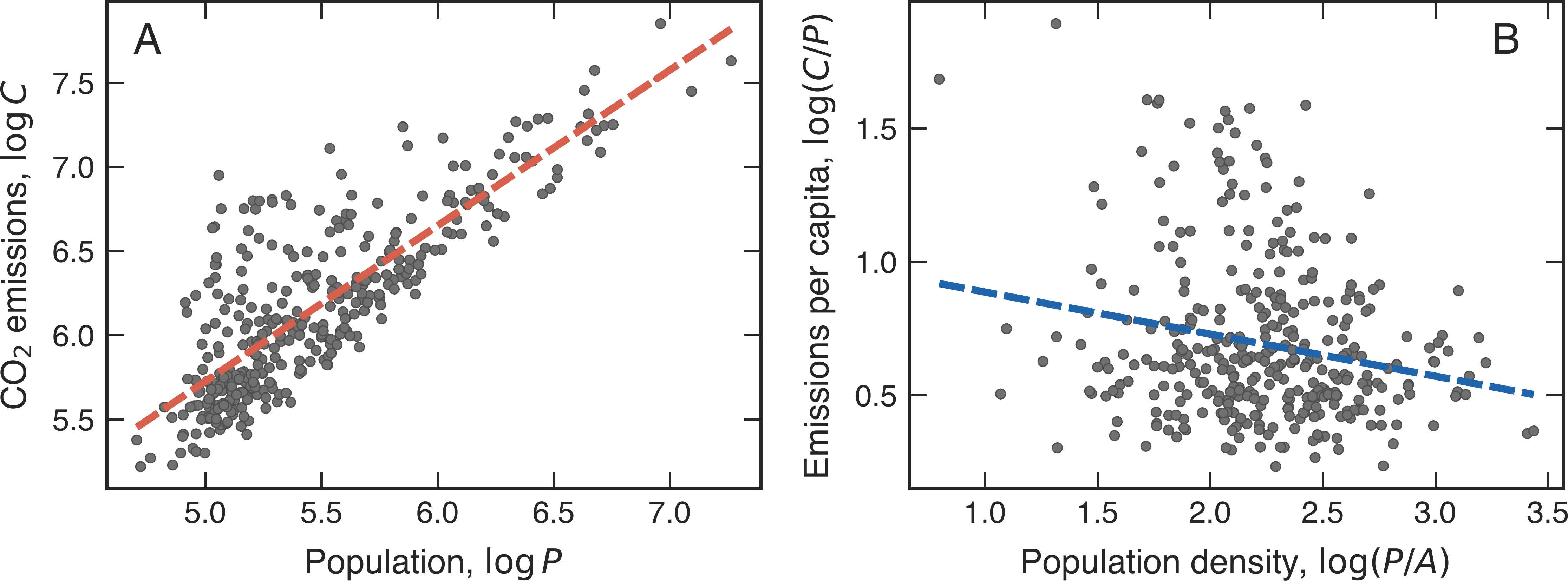}
\caption{{\bf (A) Urban scaling and (B) per capita density scaling applied to Metropolitan Statistical Areas (MSAs)}. In both plots, each dot is associated with a MSA and the dashed lines represents power-law fits [Eq.~(\ref{eq:urban_scaling})] with exponents $\beta=0.92\pm0.04$ (panel A) and $\alpha=-0.16\pm0.04$ (Panel B). Emissions data were obtained from Ref.~\cite{FragkiasLSS2013}.
}
\label{sfig:12}
\end{figure*}

\begin{figure*}[!ht]
\centering
\includegraphics[width=0.9\linewidth]{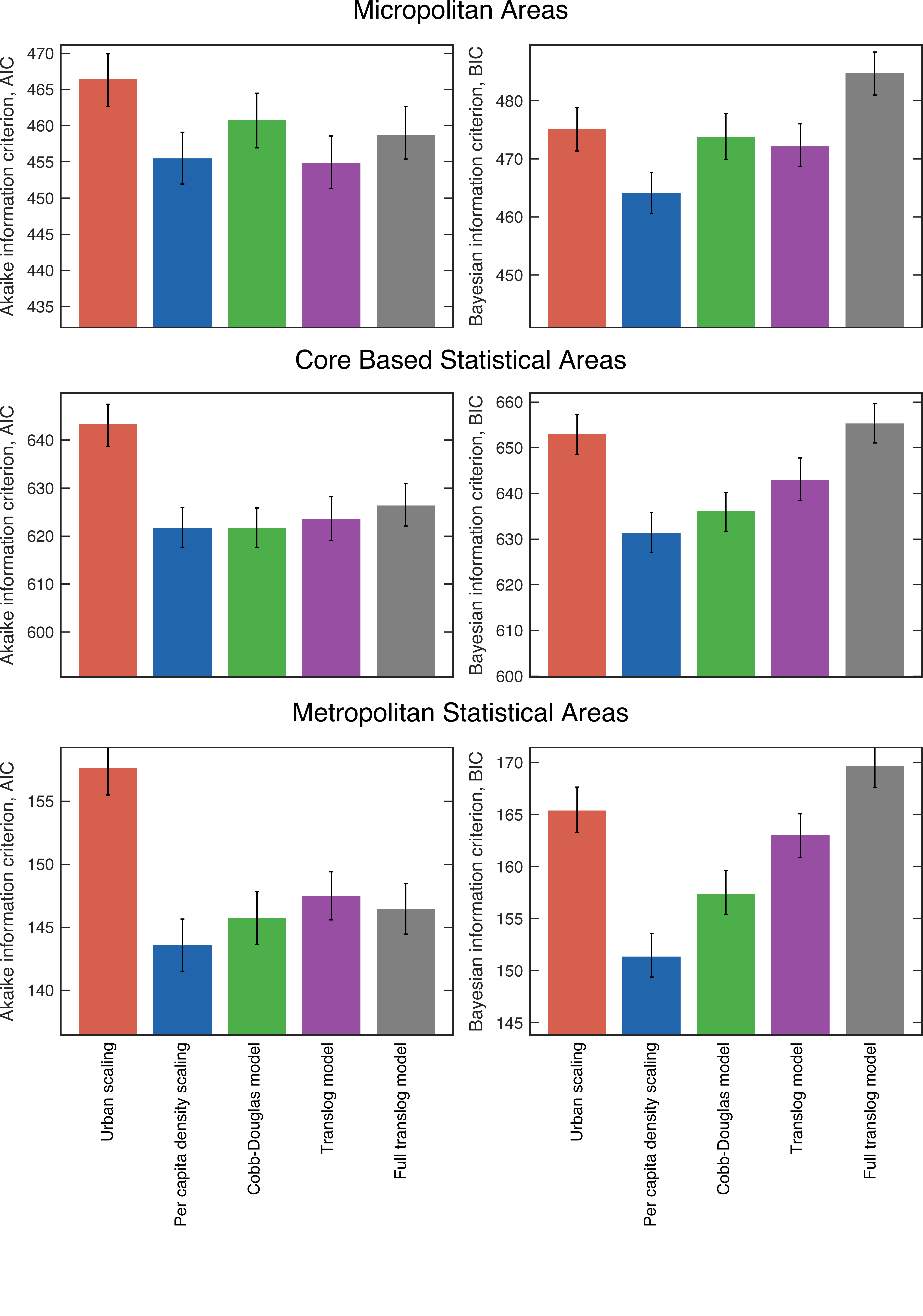}
\caption{{\bf Comparison among the exponents (A) $\beta$ and (B) $\alpha$ obtained from Metropolitan Statistical Areas (MSAs), Micropolitan Areas ($\mu$SAs), and Core Based Statistical Areas (CBSAs) data with those estimated via City Clustering Algorithm (CCA with $l=5$~km).} Bar plots in panels (A) and (B) show the average values of the parameters $\beta$ and $\alpha$ estimated after fitting the models to 1000 random samples (with replacement) of data for each definition of city. Error bars stand for the standard deviation of these values.
}
\label{sfig:13}
\end{figure*}

\begin{figure*}[!ht]
\centering
\includegraphics[width=0.8\linewidth]{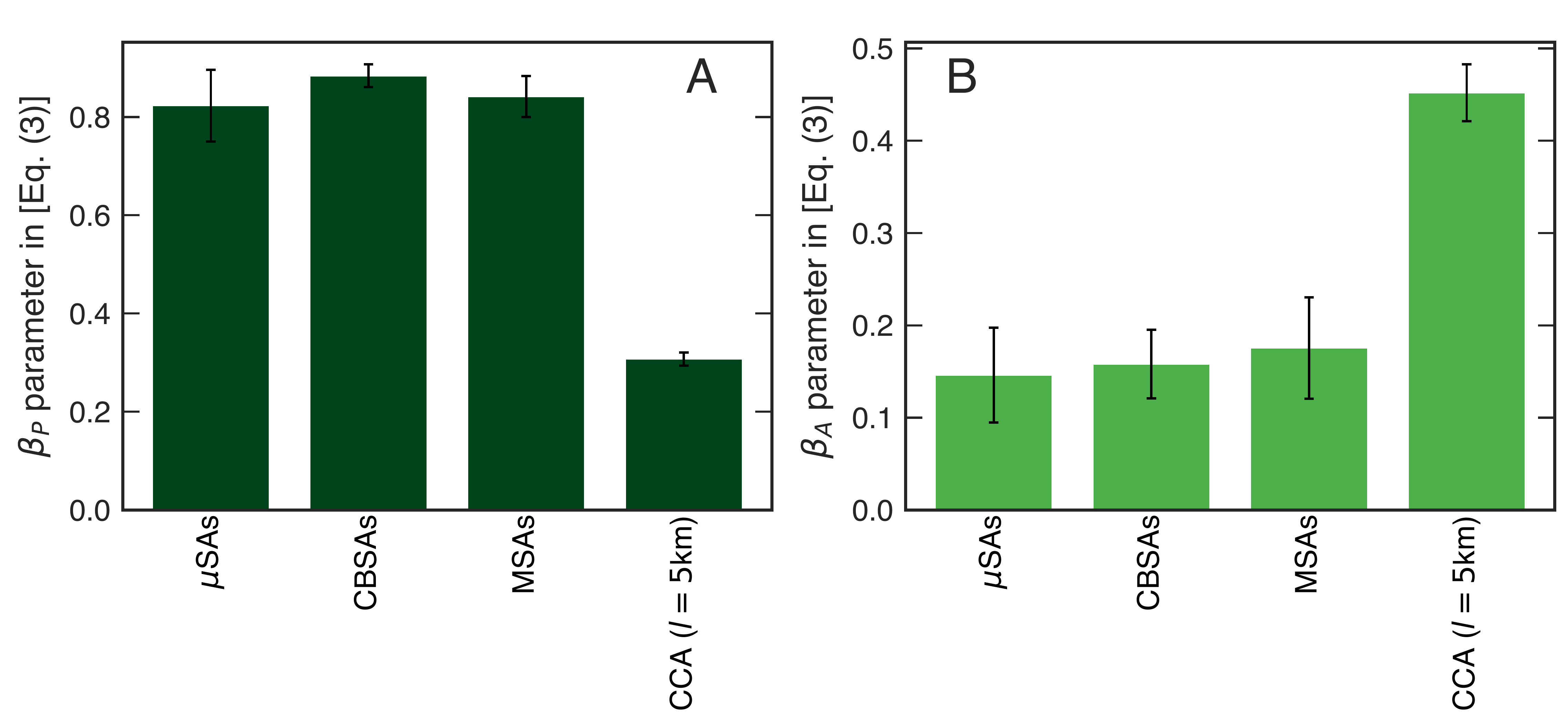}
\caption{{\bf Comparing the goodness of the fit among the models of Eqs.~(\ref{eq:urban_scaling}) [urban scaling], (\ref{eq:density_scaling}) [per capita density scaling], (\ref{eq:cobb_douglas_scaling}) [Cobb-Douglas model], (\ref{eq:translog_scaling}) [translog model], and (\ref{eq:translog_scaling_full}) [full translog model] under Metropolitan Statistical Areas (MSAs), Micropolitan Areas ($\mu$SAs), and Core Based Statistical Areas (CBSAs) data.} Each panel shows the average Akaike information criterion (AIC) and the Bayesian information criterion (BIC) estimated from 1000 different random samplings (with replacement) of data for each definition of city. In all plots, the error bars are 95\% bootstrap confidence intervals.
}
\label{sfig:14}
\end{figure*}

\begin{figure*}[!ht]
\centering
\includegraphics[width=0.9\linewidth]{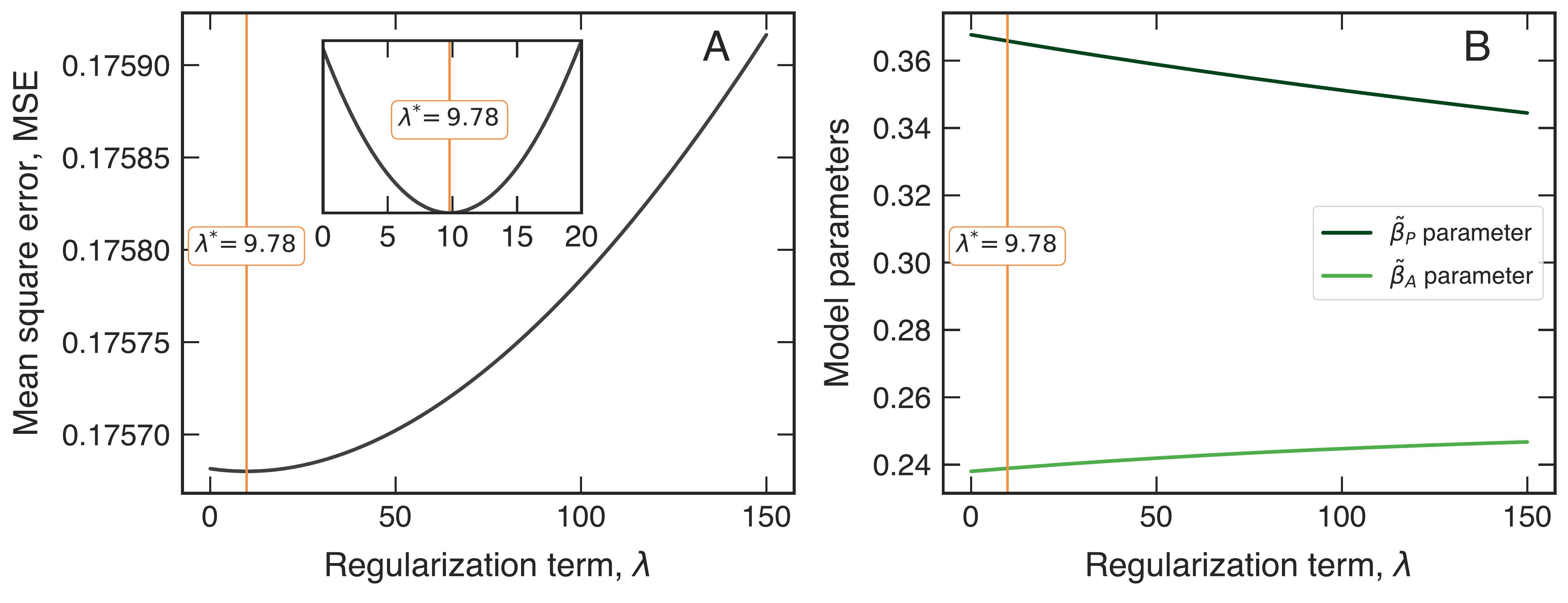}
\caption{{\bf Comparison among the Cobb-Douglas exponents [Eq.~(\ref{eq:cobb_douglas_scaling})] obtained from Metropolitan Statistical Areas (MSAs), Micropolitan Areas ($\mu$SAs), and Core Based Statistical Areas (CBSAs) data with those estimated via City Clustering Algorithm (CCA with $l=5$~km).} Bar plots in panels (A) and (B) show the average values of the parameters $\beta_P$ and $\beta_A$ estimated after fitting the models to 1000 random samples (with replacement) of data for each definition of city. Error bars stand for the standard deviation of these values.
}
\label{sfig:15}
\end{figure*}

\begin{figure*}[!ht]
\centering
\includegraphics[width=1\linewidth]{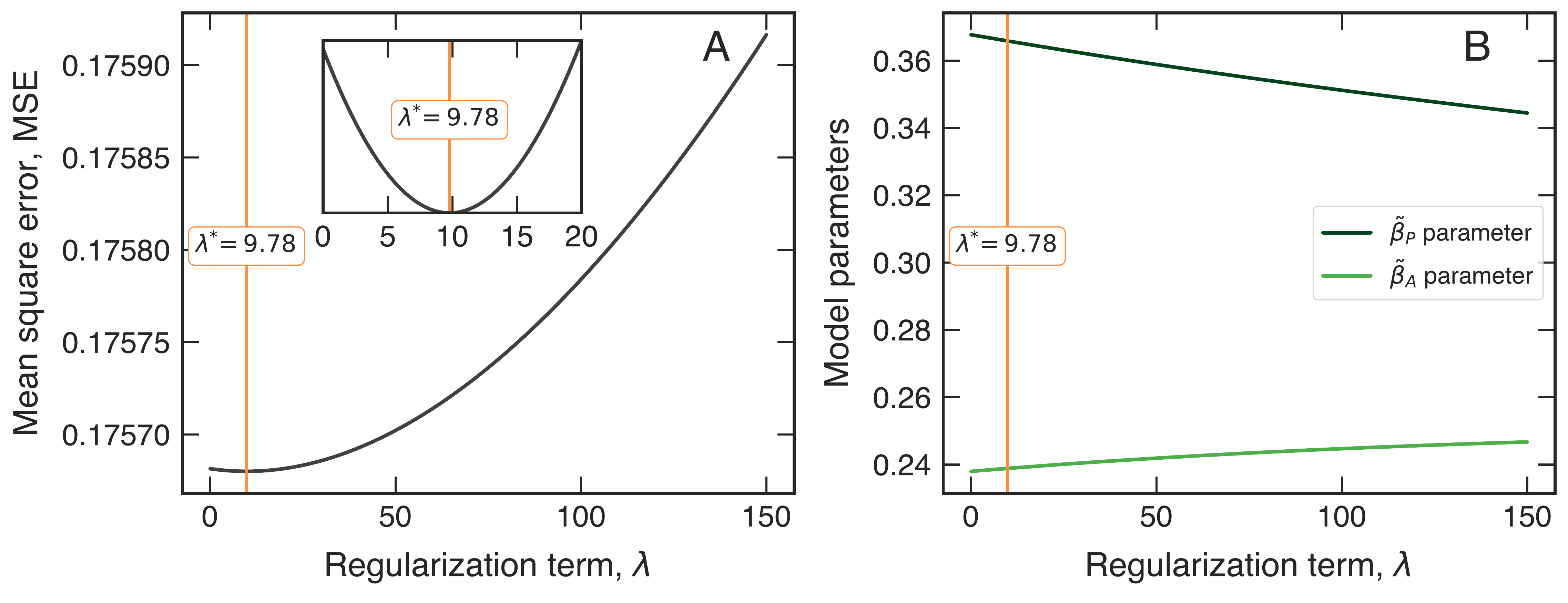}
\caption{{\bf Fitting Eq.~(\ref{eq:cobb_douglas_scaling}) [Cobb-Douglas model] to data with the ridge regression approach.} (A) The black curve shows the dependence of the mean square error (MSE) on the regularization term ($\lambda$). The average value is estimated using a leave-one-out cross validation strategy. The vertical line indicates the minimum value of the MSE occurring at $\lambda=\lambda^{*}=9.78$. The inset highlights the behavior around the minimum. (B) Dependence of the parameters $\tilde{\beta}_P$ and $\tilde{\beta}_A$ on the regularization term ($\lambda$). The optimal values for $\lambda=\lambda^{*}$ are $\tilde{\beta}_P=0.37\pm0.02$ and $\tilde{\beta}_A=0.24\pm0.02$.
}
\label{sfig:16}
\end{figure*}

\begin{figure*}[!ht]
\centering
\includegraphics[width=1\linewidth]{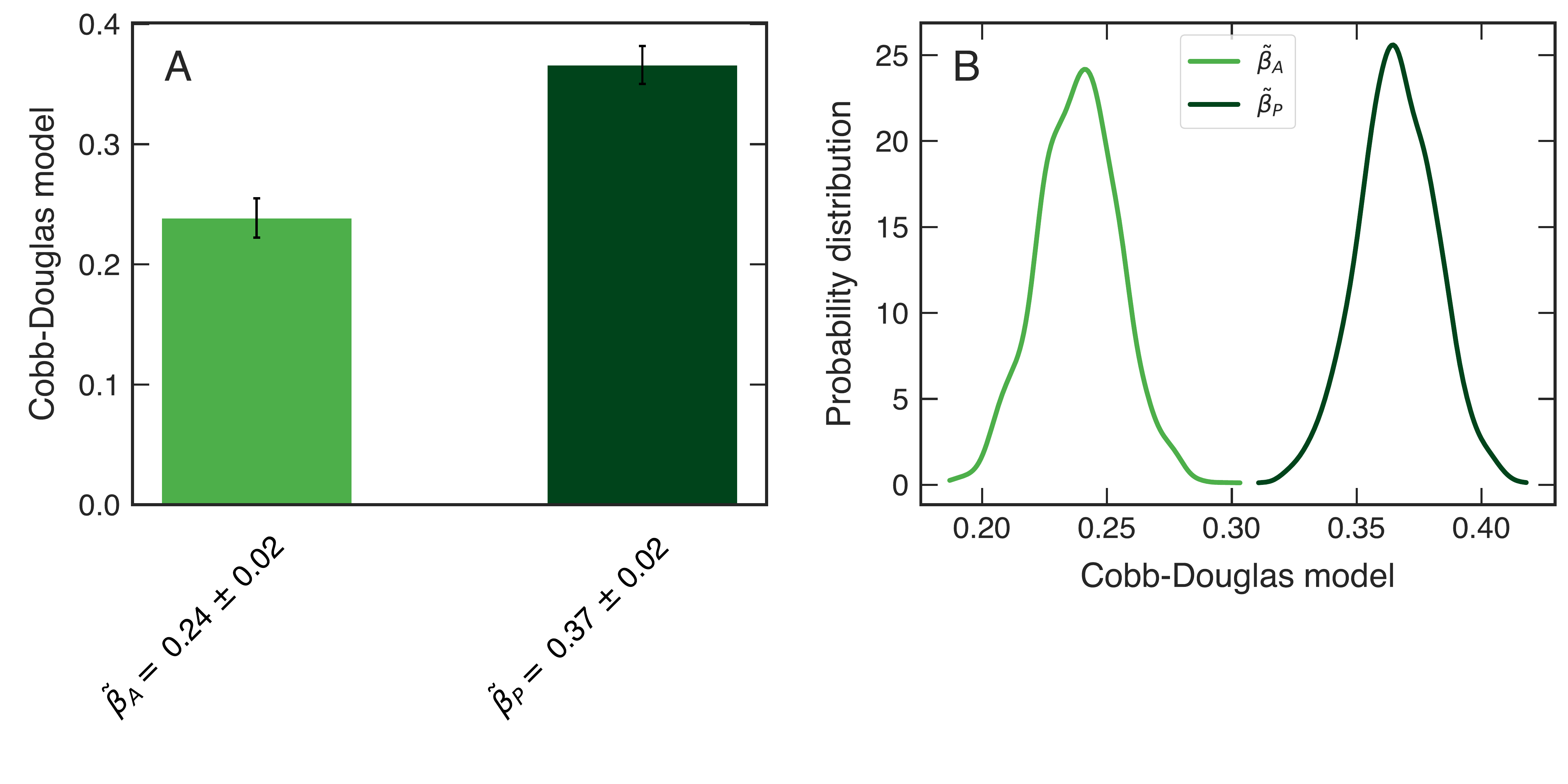}
\caption{{\bf Estimating the errors in the parameters of Eq.~(\ref{eq:cobb_douglas_scaling}) [Cobb-Douglas model].} (A) The bar plot shows the average values of the parameters $\tilde{\beta}_P$ and $\tilde{\beta}_A$ estimated after fitting the model to 1000 random samples (with replacement) of our data. Error bars stand for the standard deviation of these values. (B) Probability distribution of the values of $\tilde{\beta}_P$ and $\tilde{\beta}_A$ over all random samples. The permutation test on the model coefficients rejects the null hypothesis that they are equal to zero at the 95\% confidence level.
}
\label{sfig:17}
\end{figure*}

\begin{figure*}[!ht]
\centering
\includegraphics[width=1\linewidth]{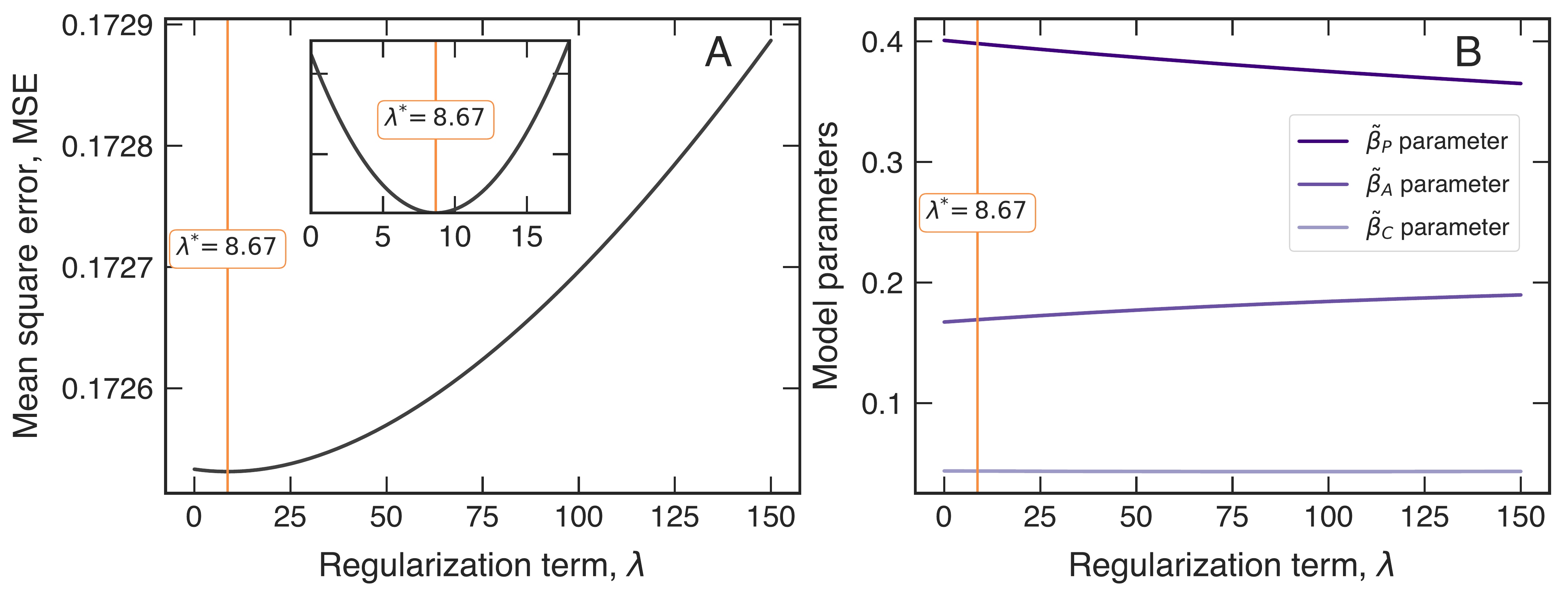}
\caption{{\bf Fitting Eq.~(\ref{eq:translog_scaling}) [translog model] to data with the ridge regression approach.} (A) The black curve shows the dependence of the mean square error (MSE) on the regularization term ($\lambda$). The average value is estimated using a leave-one-out cross validation strategy. The vertical line indicates the minimum value of the MSE occurring at $\lambda=\lambda^{*}=8.67$. The inset highlights the behavior around the minimum. (B) Dependence of the parameters $\tilde{\beta}_P$, $\tilde{\beta}_A$, and $\tilde{\beta}_C$ on the regularization term ($\lambda$). The optimal values for $\lambda=\lambda^{*}$ are $\tilde{\beta}_P=0.40\pm0.02$, $\tilde{\beta}_A=0.17\pm0.02$, and $\tilde{\beta}_C=0.044\pm0.006$.
}
\label{sfig:18}
\end{figure*}

\begin{figure*}[!ht]
\centering
\includegraphics[width=1\linewidth]{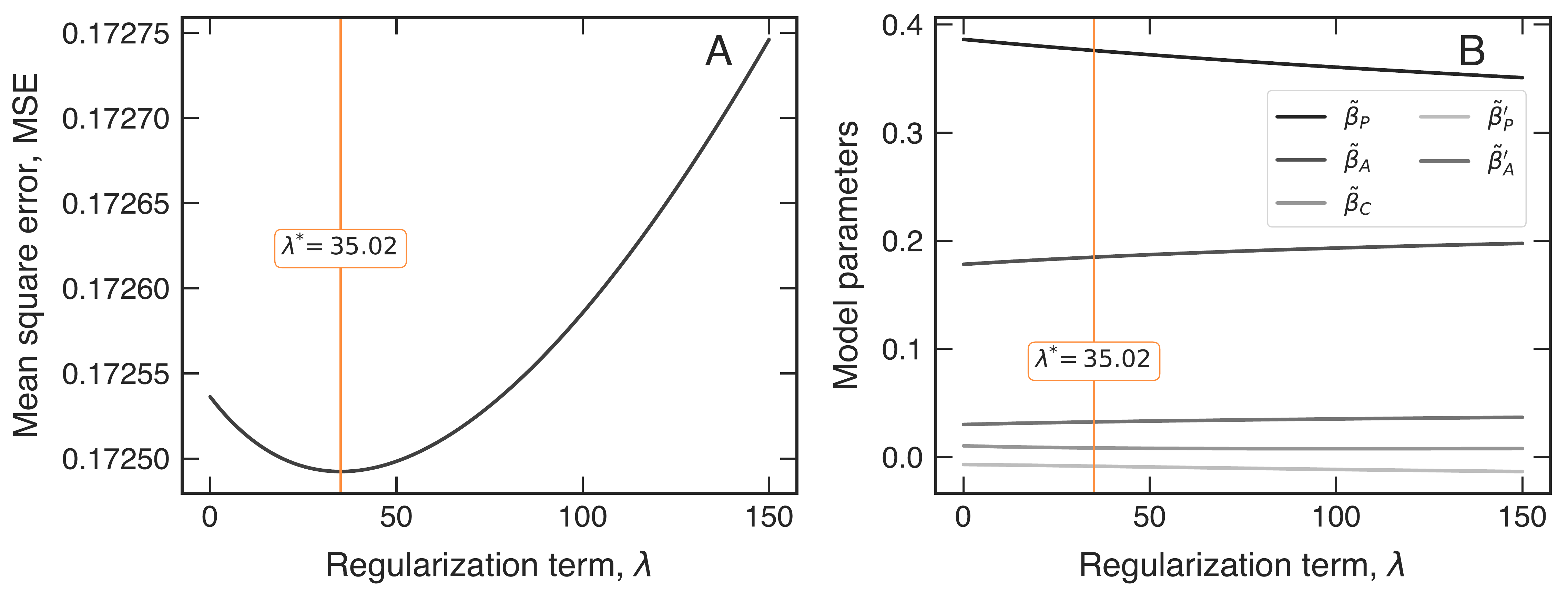}
\caption{{\bf Fitting Eq.~(\ref{eq:translog_scaling_full}) [full translog model] to data with the ridge regression approach.} (A) The black curve shows the dependence of the mean square error (MSE) on the regularization term ($\lambda$). The average value is estimated using a leave-one-out cross validation strategy. The vertical line indicates the minimum value of the MSE occurring at $\lambda=\lambda^{*}=35.02$. (B) Dependence of the parameters $\tilde{\beta}_P$, $\tilde{\beta}_A$, $\tilde{\beta}_C$, $\tilde{\beta}_P'$, and $\tilde{\beta}_A'$ on the regularization term ($\lambda$). The optimal values for $\lambda=\lambda^{*}$ are $\tilde{\beta}_P=0.378\pm0.015$, $\tilde{\beta}_A=0.184\pm0.020$, $\tilde{\beta}_C=0.009\pm0.039$, $\tilde{\beta}_P'=-0.009\pm0.021$, and $\tilde{\beta}_A'=-0.032\pm0.020$.
}
\label{sfig:19}
\end{figure*}

\begin{figure*}[!ht]
\centering
\includegraphics[width=1\linewidth]{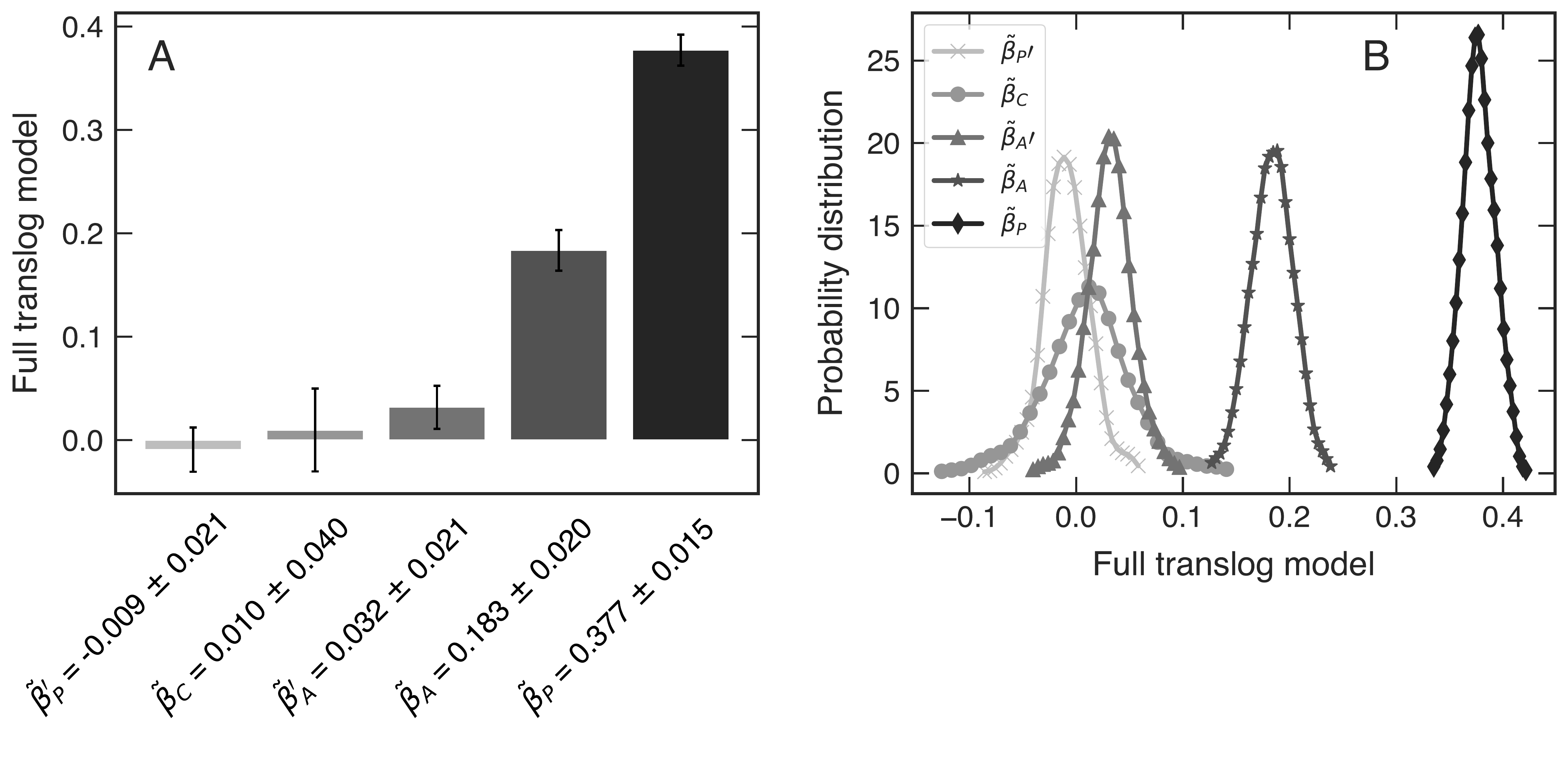}
\caption{{\bf Estimating the errors in the parameters of Eq.~(\ref{eq:translog_scaling_full}) [full translog model].} (A) The bar plot shows the average values of the parameters $\tilde{\beta}_P$, $\tilde{\beta}_A$, $\tilde{\beta}_C$, $\tilde{\beta}_P'$, and $\tilde{\beta}_A'$ estimated after fitting the model to 1000 random samples (with replacement) of our data. Error bars stand for the standard deviation of these values. (B) Probability distribution of the values of $\tilde{\beta}_P$, $\tilde{\beta}_A$, $\tilde{\beta}_C$, $\tilde{\beta}_P'$, and $\tilde{\beta}_A'$ over all random samples. The permutation test on the model coefficients rejects the null hypothesis that $\tilde{\beta}_P$, $\tilde{\beta}_A$ and $\tilde{\beta}_A'$ are equal to zero; but cannot reject the same hypothesis for $\tilde{\beta}_P'$ and $\tilde{\beta}_C$ at the 95\% confidence level.
}
\label{sfig:20}
\end{figure*}

\clearpage

\begin{table}[hb]
\vspace*{-1.3cm}
\setlength{\tabcolsep}{4pt}
\renewcommand{\arraystretch}{0.8}

\begin{footnotesize}
\begin{tabular}{ccccccccc}
\toprule
\begin{tabular}{@{}c@{}}{\bf Method,} \\{\bf reference}\end{tabular} & \textbf{Country} & \textbf{Year}& \textbf{Urban Units} & \begin{tabular}{@{}c@{}}{\bf Sample} \\{\bf Size}\end{tabular} & \textbf{Sector} & \begin{tabular}{@{}c@{}}{\bf Value} \\{\bf (Lower,Upper)}\end{tabular}& \\
\toprule
 &  &  &  & &  & $\beta$ \\
\toprule
RMA~\cite{gudipudi2019efficient} & Annex I & 2005 & Large Cities & 22 & N/A & $0.87~(0.59,0.95)$ \\
\colrule
RMA~\cite{gudipudi2019efficient} & \begin{tabular}{@{}c@{}}Non-\\Annex I\end{tabular} & 2005 & Large Cities & 39 & N/A & $1.18~(0.90,1.49)$ \\
\colrule
OLS~\cite{muller2017does} & US & \begin{tabular}{@{}c@{}}1999-\\2011\end{tabular} & \begin{tabular}{@{}c@{}}Metropolitan\\Areas\end{tabular} & 1875 & \begin{tabular}{@{}c@{}}Local CO2\\ and five air pollutants\end{tabular} & $ 0.75~(0.71,0.79)$ \\
\colrule
OLS~\cite{chang2017there} & US & \begin{tabular}{@{}c@{}}1982-\\2011\end{tabular} & Urban Centers & 101 & \begin{tabular}{@{}c@{}}Excess CO$_2$ \\from congestion\end{tabular} & $1.27~(1.21,1.33)$ \\
\colrule
PW~\cite{chang2017there} & US & \begin{tabular}{@{}c@{}}1982-\\2011\end{tabular} & Urban Centers & 101 & \begin{tabular}{@{}c@{}}Excess CO$_2$ \\from congestion\end{tabular} & $1.14~(1.38,1.44)$ \\
\colrule
OLS~\cite{rybski2017cities} & US & N/A & N/A & 122 & N/A & $ 0.92~(0.90,0.95)$ \\
\colrule
OLS~\cite{rybski2017cities} & India & N/A & N/A & 42 & N/A & $ 1.13~(1.02,1.23)$ \\
\colrule
OLS~\cite{rybski2017cities} & UK & N/A & N/A & 35 & N/A & $ 0.76~(0.68,0.83)$ \\
\colrule
OLS~\cite{mohajeri2015co} & GB & 2010 & \begin{tabular}{@{}c@{}}City \&\\Administrative \\Boundaries\end{tabular}  & 28 & \begin{tabular}{@{}c@{}}Transport \\CO$_2$ emissions\end{tabular} & $1.02~(0.96,1.08)$ \\
\colrule
OLS~\cite{LoufB2014SciRep} & US & 2010 & \begin{tabular}{@{}c@{}}Metropolitan\\Areas\end{tabular} & 101 & \begin{tabular}{@{}c@{}}Excess CO$_2$ \\from congestion\end{tabular} & $1.26~(1.17,1.35)$ \\
\colrule
OLS~\cite{LoufB2014SciRep} & \begin{tabular}{@{}c@{}}OECD\\Countries\end{tabular} & 2014 & \begin{tabular}{@{}c@{}}Metropolitan\\Areas\end{tabular} & 268 & \begin{tabular}{@{}c@{}}Transport CO$_2$ \\emissions\end{tabular} & $1.21~(1.11,1.31)$ \\
\colrule
OLS~\cite{OliveiraAM2014} & US & 2002 & \begin{tabular}{@{}c@{}}CCA clusters\\ ($l=5$~km,\\ $D=1000$)\end{tabular} & 2281 & \begin{tabular}{@{}c@{}}All Vulcan \\sectors\end{tabular} & $1.38~(1.35,1.41)$ \\
\colrule
OLS~\cite{OliveiraAM2014} & US & 2002 & \begin{tabular}{@{}c@{}}CCA clusters\\ ($l>10$~km,\\ $1000<D<4000$)\end{tabular} & N/A & \begin{tabular}{@{}c@{}}All Vulcan \\sectors\end{tabular} & $1.46~(1.44,1.48)$ \\
\colrule
OLS~\cite{FragkiasLSS2013} & US & \begin{tabular}{@{}c@{}}1999-\\2008\end{tabular} & \begin{tabular}{@{}c@{}}Metropolitan\\Areas\end{tabular}  & 933 & \begin{tabular}{@{}c@{}}All Vulcan \\sectors\end{tabular} & $ 0.93~(0.92,0.95)$ \\
\toprule
&  &  &  & &  & $\alpha$ \\
\toprule
OLS~\cite{Gudipudi2016} & US & 2000 & \begin{tabular}{@{}c@{}}CCA/GRUMP/\\GLC \\($l=1$~km)\end{tabular} & 4585 & \begin{tabular}{@{}c@{}}Buildings and \\ on-road emissions\end{tabular} & $ -0.78~(\text{N/A},\text{N/A})$ \\
\colrule
OLS~\cite{Gudipudi2016} & US & 2000 & \begin{tabular}{@{}c@{}}CCA/GRUMP/\\GLC \\($l=5$~km)\end{tabular} & 3285 & \begin{tabular}{@{}c@{}}Buildings and \\ on-road emissions\end{tabular} & $ -0.79~(\text{N/A},\text{N/A})$ \\
\colrule
OLS~\cite{Gudipudi2016} & US & 2000 & \begin{tabular}{@{}c@{}}CCA/GRUMP/\\GLC \\($l=10$~km)\end{tabular} & 2786 & \begin{tabular}{@{}c@{}}Buildings and \\ on-road emissions\end{tabular} & $ -0.82~(\text{N/A},\text{N/A})$ \\
\colrule
OLS~\cite{Gudipudi2016} & US & 2000 & \begin{tabular}{@{}c@{}}CCA/GRUMP \\($l=1$~km)\end{tabular} & 5182 & \begin{tabular}{@{}c@{}}Buildings and \\ on-road emissions\end{tabular} & $ -0.90~(\text{N/A},\text{N/A})$ \\
\colrule
OLS~\cite{Gudipudi2016} & US & 2000 & \begin{tabular}{@{}c@{}}CCA/GRUMP \\($l=5$~km)\end{tabular} & 2156 & \begin{tabular}{@{}c@{}}Buildings and \\ on-road emissions\end{tabular} & $ -1.10~(\text{N/A},\text{N/A})$ \\
\colrule
OLS~\cite{Gudipudi2016} & US & 2000 & \begin{tabular}{@{}c@{}}CCA/GRUMP \\($l=10$~km)\end{tabular} & 1538 & \begin{tabular}{@{}c@{}}Buildings and \\ on-road emissions\end{tabular} & $ -1.13~(\text{N/A},\text{N/A})$ \\
\colrule
OLS~\cite{NewmanK1989} & Global & 1980 & \begin{tabular}{@{}c@{}}World\\cities\end{tabular} & 32 & \begin{tabular}{@{}c@{}}Gasoline\\consumption\end{tabular} & $ -0.92~(-0.85,-0.99)$ \\
\colrule
OLS~\cite{baur2013urban} & EU & 2007-2009 & \begin{tabular}{@{}c@{}}European\\cities\end{tabular} & 134 & \begin{tabular}{@{}c@{}}Transport\\CO$_2$ emissions\end{tabular} & $ -0.19~(\text{N/A},\text{N/A})$ \\
\colrule
OLS~\cite{baur2013urban} & EU & 2007-2009 & \begin{tabular}{@{}c@{}}France\end{tabular} & 24 & \begin{tabular}{@{}c@{}}Transport\\CO$_2$ emissions\end{tabular} & $ -0.39~(\text{N/A},\text{N/A})$ \\
\colrule
OLS~\cite{baur2013urban} & EU & 2007-2009 & \begin{tabular}{@{}c@{}}Spain\end{tabular} & 22 & \begin{tabular}{@{}c@{}}Transport\\CO$_2$ emissions\end{tabular} & $ -0.40~(\text{N/A},\text{N/A})$ \\
\botrule

\end{tabular}
\end{footnotesize}
\caption{Exponents associated with the urban scaling ($\beta$) and per capita density scaling ($\alpha$). Abbreviations: Reduced Major Axis Regression (RMA), Ordinary Least Square Regression (OLS), Prais-Winsten Regression (PW).}\label{stab:1}
\end{table}

\begin{table}[hb]
\setlength{\tabcolsep}{4pt}
\renewcommand{\arraystretch}{0.8}

\begin{footnotesize}
\begin{tabular}{ccc}
\toprule
{\bf Property} & {\bf Cobb-Douglas model, Eq.~(\ref{eq:cobb_douglas_scaling})} & {\bf Translog model, Eq.~(\ref{eq:translog_scaling})} \\
\toprule
\begin{tabular}{@{}c@{}} Elasticity of\\scale in terms\\of $P$ and $A$ \end{tabular} & $\beta_P+\beta_A$ & $\beta_P+\beta_A + \beta_C \log(P A)$ \\
\colrule
\begin{tabular}{@{}c@{}} Elasticity of\\scale in terms\\of $P$ and $P/A$ \end{tabular} & $\beta_P$ & $\beta_P + \beta_C \log A$\\
\colrule
\begin{tabular}{@{}c@{}} Technical rate\\of substitution\\between $P$ and $A$ \end{tabular} & $-\frac{\beta_P}{\beta_A(P/A)}$ & $\frac{-1}{(P/A)}\left(\frac{\beta_P +\beta_C \log A}{\beta_A +\beta_C \log P}\right)$\\
\colrule
\begin{tabular}{@{}c@{}} Elasticity of\\substitution \end{tabular} & 1 & $- \frac{\beta_P+\beta_A + \beta_C \log(P A)}{C(\beta_P+\beta_A + \beta_C \log(P A) -2\beta_C)}$\\
\colrule
\begin{tabular}{@{}c@{}} Marginal product\\of population\\(in terms of $P$ and $A$) \end{tabular} & $\beta_P$ & $\beta_P +\beta_C \log A$\\
\colrule
\begin{tabular}{@{}c@{}} Marginal product\\of area\\(in terms of $P$ and $A$) \end{tabular} & $\beta_A$ & $\beta_A +\beta_C \log P$\\
\colrule
\begin{tabular}{@{}c@{}} Marginal product\\of population\\(in terms of $P$ and $P/A$) \end{tabular} & $\beta_P +\beta_A$ & $\beta_P +\beta_A +\beta_C\log(PA)$\\
\colrule
\begin{tabular}{@{}c@{}} Marginal product\\of density\\(in terms of $P$ and $P/A$) \end{tabular} & $-\beta_A$ & $-\beta_A -\beta_C \log P$\\
\colrule
\begin{tabular}{@{}c@{}} Condition for\\population dominates\\over density \end{tabular} & \begin{tabular}{@{}c@{}}$|\beta_P+\beta_A|>|\beta_A|$ \end{tabular} & \begin{tabular}{@{}c@{}}$A> 10^{-\beta_P/\beta_A}$ \\ (if $\beta_P,\beta_A,\beta_C>0$ and $P,A>1$)\end{tabular}\\
\colrule
\begin{tabular}{@{}c@{}} Condition for\\decreasing returns\\to scale\\(in terms of $P$ and $A$) \end{tabular} & $\beta_P+\beta_A<0$ & $P A < 10^{\frac{1-\beta_P-\beta_A}{\beta_C}}$\\
\colrule
\begin{tabular}{@{}c@{}} Condition for\\decreasing returns\\to scale\\(in terms of $P$ and $P/A$) \end{tabular} & $\beta_P<0$ & $A < 10^{\frac{1-\beta_P}{\beta_P}}$\\
\botrule

\end{tabular}
\end{footnotesize}
\caption{Summary of the main properties of Cobb-Douglas [Eq.~(\ref{eq:cobb_douglas_scaling})] and translog [Eq.~(\ref{eq:translog_scaling})] models. All marginal products are expressed in logarithmic scale.}\label{stab:2}
\end{table}

\end{document}